\documentclass[letterpaper,twocolumn,10pt]{article}
\usepackage{usenix}

\usepackage[utf8]{inputenc}
\usepackage[english]{babel}
\usepackage{blindtext}
\usepackage{textcomp}
\usepackage{amsmath,amssymb,amsfonts}
\usepackage{makecell}
\usepackage{graphicx}
\usepackage{comment}
\usepackage{multirow}
\usepackage{fancyhdr}
\usepackage{xspace}
\usepackage{xcolor}
\usepackage{dsfont}
\usepackage{pgfplots}
\usepackage{multirow}
\usepackage{float}
\usepackage{microtype}
\usepackage{subfig}
\usepackage{booktabs}
\usepackage{hyperref}
\usepackage{algorithmic}
\usepackage{bussproofs}
\usepackage{mathtools}

\mathtoolsset{centercolon}

\pgfplotsset{compat=1.17}

\def \OurSystem{Rhino\xspace}

\title{Auto-Parallelizing Large Models with \OurSystem: \\A Systematic Approach on Production AI Platform}

\newcommand*\samethanks[1][\value{footnote}]{\footnotemark[#1]}
\author{
{\rm Shiwei Zhang}$^{1,2,}$\thanks{Shiwei and Lansong contributed equally.} \qquad
{\rm Lansong Diao}$^{1,}$\samethanks \qquad
{\rm Siyu Wang}$^1$ \qquad
{\rm Zongyan Cao}$^1$ \qquad
{\rm Yiliang Gu}$^1$ \\ \qquad
{\rm Chang Si}$^1$ \qquad
{\rm Ziji Shi}$^{1,3}$ \qquad
{\rm Zhen Zheng}$^1$ \qquad
{\rm Chuan Wu}$^2$ \qquad
{\rm Wei Lin}$^1$ \\ \\
$^1$Alibaba Group \qquad
$^2$The University of Hong Kong \qquad
$^3$National University of Singapore \\
}

\begin{document}

\maketitle

\begin{abstract}
    We present \OurSystem, a system for accelerating tensor programs with automatic parallelization on AI platform for
    real production environment. It transforms a tensor program written for a single device into an equivalent
    distributed program that is capable of scaling up to thousands of devices with no user configuration. \OurSystem
    firstly works on a semantically independent intermediate representation of tensor programs, which facilitates its
    generalization to unprecedented applications. Additionally, it implements a task-oriented controller and a distributed
    runtime for optimal performance. \OurSystem explores on a complete and systematic parallelization strategy space
    that comprises all the paradigms commonly employed in deep learning (DL), in addition to strided partitioning and
    pipeline parallelism on non-linear models. Aiming to efficiently search for a near-optimal parallel execution plan,
    our analysis of production clusters reveals general heuristics to speed up the strategy search. On top of it, two
    optimization levels are designed to offer users flexible trade-offs between the search time and strategy quality.
    Our experiments demonstrate that \OurSystem can not only re-discover the expert-crafted strategies of classic,
    research and production DL models, but also identify novel parallelization strategies which surpass existing systems
    for novel models.
\end{abstract}

\section{Introduction}

Large models with hundreds of billions of parameters have exhibited unprecedented performance and driven
the research of new deep neural networks (DNNs). For example, PaLM \cite{palm} for language modeling has 540B
parameters and M6 \cite{m6} for multimodal pretraining contains 100B parameters. Training such large scale models on
commodity hardware (e.g. GPUs) requires carefully designed distributed training strategies that often employ various
parallelism paradigms including data parallelism, model parallelism, and pipeline parallelism.

Many systems have explored automating the parallelization of large models. Tofu \cite{tofu} finds hybrid data and model
parallelism strategies with dynamic programming algorithm. Alpa \cite{alpa} automates intra-operator parallelism with
Integer Linear Programming (ILP) and inter-operator parallelism with dynamic programming. Unity \cite{unity} supports
joint optimization of algebraic transformations and parallelization by randomized graph substitution on a parallel
computation graph. These systems show that fully automated algorithms can achieve comparable or superior distributed
training performance than expert-designed strategies.

However, state-of-the-art models, particularly those large models that necessitate distributed training over large clusters, still
tend to employ manually designed parallelization strategies. MT-NLG \cite{mtnlg} is trained using manually designed 3D
parallelism (hybrid data parallelism, model parallelism, and pipeline parallelism) on DeepSpeed \cite{deepspeed} and
Megatron \cite{megatron}. GLaM \cite{glam} is trained with manually designed SPMD parallelism (hybrid data parallelism
and model parallelism) using GSPMD \cite{gspmd}. This prompts us to ponder: What impediments exist in the current auto-parallelization frameworks that preventing them from being used in production?

Our arguments can be summarized in the following four aspects.

First, existing auto-parallelization systems are not general enough as they often design the distributed training
strategies based on the semantics of a DNN model, such as the concept of ``layers'', the meaning of a tensor axis
(``batchsize'', ``tokens'', or ``hidden width''), etc. For example, AccPar \cite{accpar} considers three types of
parallelism based on the semantics of tensor dimensions. PipeDream \cite{pipedream} partitions the models on the
``layer'' level. Whale \cite{whale} leverages manual annotations of layer groups to achieve pipeline parallelism.
However, as new DNN architectures are being proposed, the notions of tailored layers may become obsolete and prevent the
opportunity for novel parallel strategies. For example, Tofu \cite{tofu} utilizes high-level programming API provided in DNN
frameworks to detect and merge unrolled timesteps for RNN layers; this method will fail for new models that do not
contain RNN layers (like the Transformer-based models). Additionally, while tying the distributed training strategies with
model semantics facilitates reasoning and debugging, it restricts the strategy space and may miss important optimization
opportunities. As an example, a layer, often mapped to a \texttt{nn.Module} object in PyTorch, is a semantical group of
operators that are often chosen based on the code reusability. Using layers as the partitioning unit unnecessarily
couples the programming convention with model training performance.

Second, many auto-parallelization systems are limited in the range of their parallelism paradigms and do not provide complete and systematic
parallelization method space. For intance, PipeDream \cite{pipedream} fails to support tensor parallelism and Tofu
\cite{tofu} does not consider pipeline parallelism. In addition, no existing SPMD systems currently support strided partitioning that will be explained in Sec.~\ref{sec:sharding_spec}. With the advent of ever larger models, it has become essential to combine all known parallelism paradigms to maximize the parallelization
degree.

Third, the heuristics used in existing auto-parallelization systems are based largely on intuition and only evaluated in
very limited scenarios. Since the space of parallelization strategies is exponentially large, all existing
auto-parallelization systems employ heuristics to reduce the search space and accelerate the searching.
These heuristics significantly influence the quality of the generated parallelization strategy, yet are mainly conceived out of intuition and rarely verified with the diverse workloads for robustness.

Finally, many auto-parallelization systems are near black-box and not intervenable. For example, it may be unclear how a
user can resolve the problem if the auto-parallelization system outputs a strategy that exceeds the GPU memory limit and
leads to an out-of-memory(OOM) error. Hence, a reliable auto-parallelization system ought to enable sufficient control to handle unforeseen situations.

Addressing these issues, we present \OurSystem, an  fully automatic parallelization training system designed with the following
guiding principles:

$\triangleright$ \OurSystem works on HLO\cite{jax} level, the compiler intermediate representation (IR) used by XLA
\cite{xla}. The HLO IR contains the full computation graph including forward and backward passes, enabling holistic
optimization and allowing potentially different sharding strategies on the forward and backward of the same layer.
Further, \OurSystem does not rely on the notion of layers or the roles of tensor axes, but rather focuses solely on the
arithmetic nature of the operation sequence, thus allowing decoupling of the model declaration code (different front-end
libraries like Tensorflow or JAX, different layer boundaries, etc.) and its training performance, as long as the
programs result in the same computation. Moreover, \OurSystem supports new training methods other than backpropagation,
such as the forward-forward algorithm \cite{forwardforward}.

$\triangleright$ \OurSystem contains a complete and systematic strategy space. Apart from the sharding dimensions,
\OurSystem also considers the strided layout of tensor shards, which has revealed useful sharding strategies that have
not been previously reported. We demonstrate that many common parallelization strategies, such as Megatron's
\cite{megatron} model parallelism, ZeRO optimizer \cite{zero}, and GShard's \cite{gshard} MoE sharding strategy, are
special cases of the strategies explored in \OurSystem. To efficiently explore the large search space, we investigate
the common patterns in production DNN models and propose a search algorithm that combines integer linear programming
(ILP) and dynamic programming (DP).


$\triangleright$ \OurSystem uses data-driven and systematically evaluated heuristics derived from a production system to
expedite the strategy search. With the adequate cluster task information collected online, we construct the precise and
refined statistics concerning systematic optimization capabilities, thus improving \OurSystem and enabling its
deployment in production. Considering that it is impractical to conduct an exhaustive search on the exponentially large
strategy space at the HLO level, \OurSystem incorporates the domain knowledge and insights gathered through analyzing
production models to achieve fast and reliable strategy search. 

$\triangleright$ \OurSystem supports full automatic parallelization but also allows for adequate user intervention, which employs user annotation for tuning the partition strategy in a semi-automatic way.
Further, \OurSystem provides two optimization levels to trade-off between search time and strategy quality for different
scenarios. Finally, our practice enhances the capability of corner cases handling, thereby further improving \OurSystem.

For implementation, \OurSystem develops task graph and employs static scheduling to achieve better throughput by
reducing collaborative overhead that will be explained in Sec.~\ref{sec:implementation}. Our experiments show that
\OurSystem can achieve better throughput for 19\% \textasciitilde 25\% when compared with state-of-the-art systems. For
large models, \OurSystem includes an aggressive optimization level that speeds up the strategy search by 1.5\texttimes{}
\textasciitilde 3\texttimes{}, with only 10\% performance cost. Based on these production-friendly advantages,
\OurSystem supports multiple models training on production AI platform.

\section{Background}

\subsection{Common Patterns in Distributed DNN Training}

\textbf{SPMD parallelism.} SPMD parallelism \cite{gspmd,alpa} generalizes data parallelism and model parallelism, allowing each to be evenly partitioned along any of its dimensions and distributed across different devices. 
When the partition methods of two operators are incompatible, i.e. they produce and consume tensors with different
sharding methods, MPI-style \cite{mpi} collective communication is necessary to split and merge the tensor shards. Common
collective communication operators include \texttt{All-Reduce}, \texttt{All-Gather}, \texttt{Reduce-Scatter}, and
\texttt{All-To-All}.

\textbf{Pipeline parallelism.} With pipeline parallelism \cite{dapple,pipedream,gpipe}, the model is
divided into a sequence of stages, which distributed across different groups of devices. The input data batch is partitioned into
microbatches and the execution of these microbatches is orchestrated in a pipelined manner. Point-to-point communication of
the activations and their gradients occur between stages. Of critical importance is the stage division and scheduling, as these have a direct impact on the performance of pipeline parallelism, and are nontrivial to optimally decide. Thus, an ideal pipeline strategy should balance the computation time of each stage, while at the same time minimizing the transfers among the stages.

Researchers have been exploring the combination of SPMD parallelism and pipeline parallelism for training large models
\cite{deepspeed,alpa,unity}. However, designing combined strategies manually requires extensive expertise and is difficult to
generalize, while automated search has to grapple with a daunting exploration space in the context of large models and complex clusters, which spends hours of compilation time for large models. 
As such, a promising
approach towards efficient and generalizable parallelization strategy search lies in guiding automated search with heuristics that
encode the expert knowledge.

\subsection{Computation Graph and HLO}

Backpropagation is the dominant training method for deep learning. AI consists of two components: a forward step, which calculates the training loss, and a backward step, which computes the gradients of the model parameters. Both of these steps can be expressed in the form of a tensor program, a sequence of operators with tensors as inputs and outputs. Many popular deep learning frameworks \cite{tensorflow,pytorch} encode tensor programs as directed acyclic graphs (DAGs) called computation graphs. Each edge in a computation graph represents a tensor, while vertices represent operations on those tensors.

Different deep learning frameworks have distinct operator sets and may be optimized for different hardware (such as TPU or GPU).
To avoid redundant engineering efforts when supporting various front-end platforms and back-end hardwares, AI
compilers (such as XLA \cite{xla} and TVM \cite{tvm}) usually introduce platform- and hardware- agnostic intermediate
representations (IRs) for tensor programs. AI compilers convert models defined using different front-end platforms into
unified computation graphs represented by the IR, followed by a sequence of optimizations passes. Each of these passes
applies a specific optimization (e.g. as dead code elimination) and transforms the given IR into an optimized one. The
optimized IR is then lowered into a low-level IR for hardware-specific optimization and compilation.

\subsection{Recent models}

We present three recently proposed models that feature both large size and intricate architectures, highlighting the complexities posed by distributed training. Through rigorous examination of the training process, we evaluate the corresponding challenges and explore the potential opportunities in the training of these models.

\textbf{Mixture-of-experts (MoE) models.} MoE models have achieved remarkable success in natural language modeling
\cite{moe,glam,switchtransformer}. MoE models comprise sparse layers (MoE layers) containing conditionally
activated experts. The routing layer decides which experts will process specific data samples. Since MoE layers tend to be extremely large, sharding of multiple dimensions often becomes necessary.

\textbf{Diffusion models.} Diffusion models employ a sequence of denoising autoencoders for image synthesis. Stable
diffusion \cite{stablediffusion} integrates UNet and attention mechanisms to support image generation that is conditioned upon
inputs like text. The model architecture contains cross-layer connections, which makes the pipeline design more intricate.

\textbf{DNABERT.} DNABERT \cite{dnabert} is a variant of the BERT \cite{bert} model tuned for genome-related tasks like
prediction of promoters, splice sites and transcription factor binding sites. Compared to the BERT models used for NLP
tasks, DNABERT has a longer sequence length in order to capture longer nucleotide contexts. This motivates parallelization
on the token dimension when training this model, which is rarely explored in existing studies.

\subsection{Production models}

We list some models used in our production cluster. These models are widely used and their training performance is
paramount for many services. In particular, for neural language processing (NLP) tasks, we employ GPT \cite{gpt3}, T5
\cite{t5}, BERT \cite{bert}, ALBERT \cite{albert}, and RoBERTa \cite{roberta}, which are comprised in an internal NLP toolkit. When it comes to computer vision (CV) related tasks,
VGG19 \cite{vgg} and ResNet \cite{resnet} are what we use on large scale facial detection. Additionally, we employ M6 \cite{m6} for multimodal tasks in recommendation.

\section{Overview}

\begin{figure}[t]
    \centering
    \includegraphics[width=0.4\textwidth]{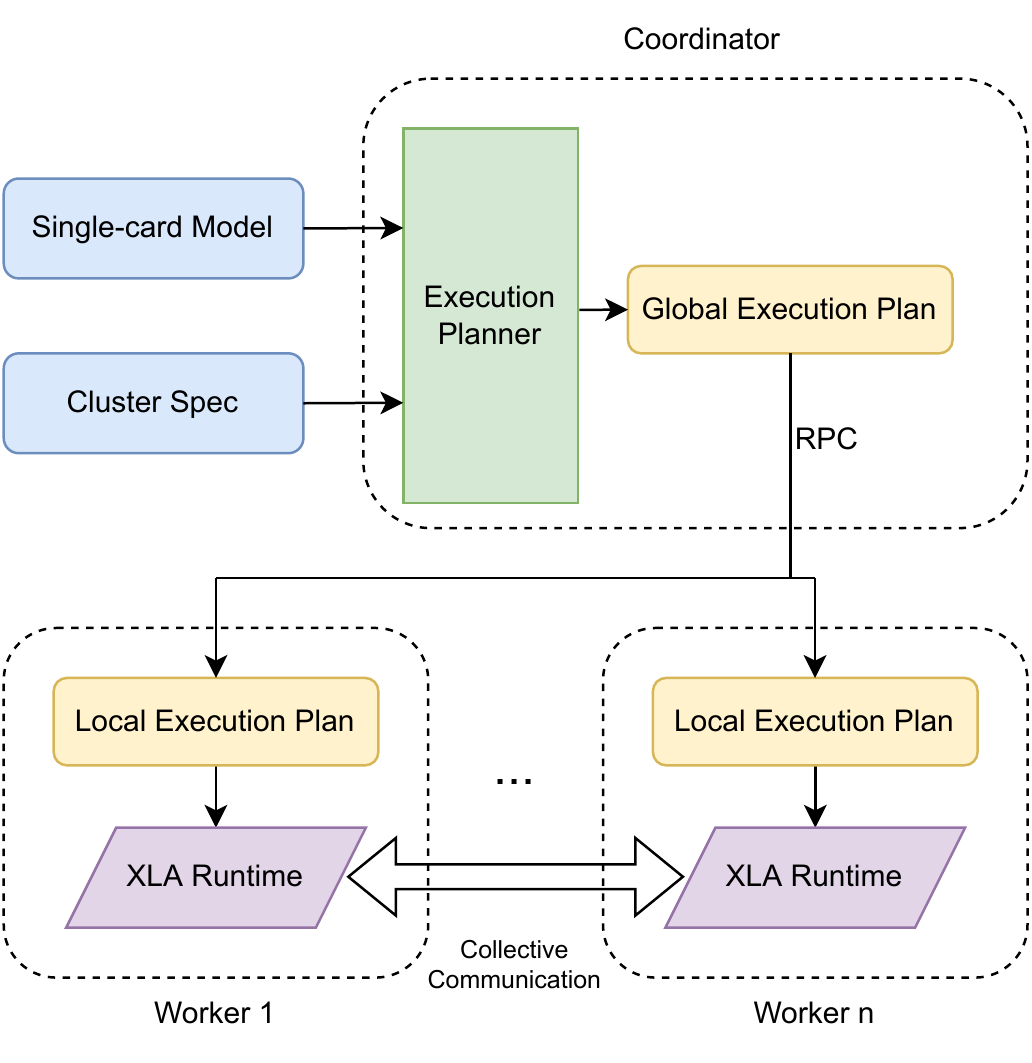}
    \caption{\OurSystem Overview\label{fig:overview}}
\end{figure}

\OurSystem is a standalone distributed training system designed for industrial production, including an execution planner and distributed runtime. As depicted in 
Fig.~\ref{fig:overview}, the execution planner first takes a single-card computation graph (in HLO IR) and cluster configuration as inputs, then optimizes and distributes the model, generating the SPMD strategy (mentioned in Sec.~\ref{sec:spmd}), pipeline strategy (explained in Sec.~\ref{sec:pipeline}), and a static
schedule (presented in Sec.~\ref{sec:static_scheduling}) plan on each device. Each device then further fetches its local
execution plan from the execution planner through RPC and utilizes \OurSystem's runtime to carry out its part in the distributed program, with very low overhead thanks to the ahead-of-time static
scheduling established over the task graph mechanism (Sec.~\ref{sec:task_graph}).

\begin{figure}[t]
    \centering
    \includegraphics[width=0.28\textwidth]{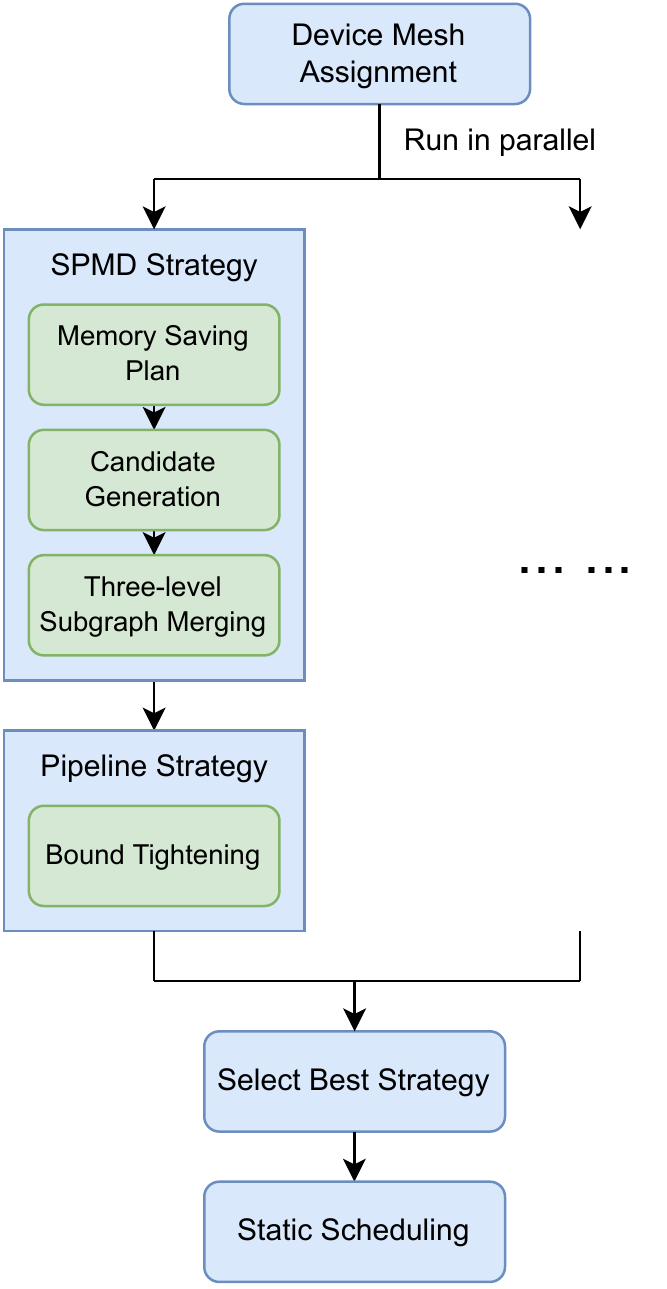}
    \caption{\OurSystem execution planner workflow.\label{fig:workflow}}
\end{figure}

The workflow of \OurSystem execution planner is illustrated in Fig.~\ref{fig:workflow}. We enumerate different device mesh 
configurations and hunt for a strategy for each configuration in parallel, through our SPMD strategy search algorithm and
pipeline planning algorithm. Subsequently, we select the optimal strategy out of different configurations and execute our static scheduling.

\OurSystem targets industrial production, hence both the compilation time and runtime performance are taken into consideration as the goal of system design. Typically, The search for both SPMD and pipeline strategies is unacceptable to the industry for an HLO model comprising quite a large number of instructions. This requires the proper design for exploring automatic parallelizations. Inspired by the feedback of deep models online, we summarize and propose a novel heuristic searching algorithm to significantly reduce the search space and accelerate the compilation with only slight sacrificing to the quality of generated parallelization strategy. For the SPMD strategy, \OurSystem employs graph coarsening and a three-level subgraph merging algorithm to achieve acceptable compilation time with the help of identifying the critical nodes and cone structure from the original graph. Then it formulates strategy exploration in each subgraph as Integer Linear Programming problem and optimizes all subgraphs together using Dynamic Programming. For pipeline strategy, \OurSystem segments the entire model through identified computationally-intensive operations without linearizing the topology order. Unlike Alpa\cite{alpa}, \OurSystem also formulates pipeline exploration as Integer Linear Programming problem based on a heuristic with a bound-tightening algorithm, which significantly facilitates the search time. Finally, we develop a central controller that makes decisions among the SPMD strategy, pipeline strategy, or hybrid strategy. All heuristics are data-driven designed and necessary parts to expedite compilation time, which contributes to handling the corner cases compared to the rule-based method.

In this paper, we explain our SPMD strategy search algorithm in Sec.~\ref{sec:spmd} and our
pipeline strategy in Sec.~\ref{sec:pipeline}. We then elaborate on the runtime implementation of \OurSystem in
Sec.~\ref{sec:implementation}. We evaluate the performance of \OurSystem in Sec.~\ref{sec:evaluation}, discuss related work in
Sec.~\ref{sec:related_work}, and finally conclude in Sec.~\ref{sec:conclusion}.

\section{SPMD Strategy\label{sec:spmd}}

To support large DNN models where a single tensor may exceed the memory limit of a GPU, intra-operator partitioning is
necessary. \OurSystem adopts the SPMD parallelism, which enables data and model parallelism. The main challenge when using 
SPMD parallelism lies in deciding the dimensions of each tensor to be partitioned so as to minimize the
overall communication time.

SPMD sharding can be repeated on orthogonal subgroups of devices. For example, the batch size dimension may be partitioned along at the machine level, while the hidden size can be partitioned at the level of the GPUs inside a machine.
In \OurSystem, the devices are organized as device meshes and the SPMD sharding is performed recursively on each
dimension of the device mesh. We enumerate different device mesh configurations, run strategy search algorithms for them
in parallel, and pick out the best strategy.

\subsection{Tensor Sharding Specs\label{sec:sharding_spec}}

We introduce the various possible sharding methods for a tensor considered in \OurSystem. A tensor is a multi-dimensional array.
For example, the input to an image classification model is usually a three-dimensional tensor, whose first and
second dimensions are the width and height of the image, and the third dimension is the color channels.

\begin{figure}[t]
    \centering
    \includegraphics[width=.36\textwidth]{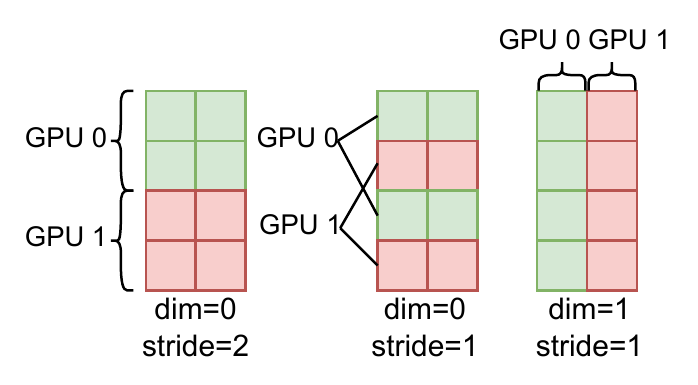}
    \caption{Different sharding methods of a 4x2 tensor.\label{fig:spmd_stride}}
\end{figure}

A tensor can be sharded in multiple ways for distributed training, called \textit{sharding spec}.
Fig.~\ref{fig:spmd_stride} illustrates different sharding spec for a 4x2 tensor. In \OurSystem, the sharding spec is
denoted by a tuple $(dim, stride)$, where $dim$ defines the dimension of a tensor to be partitioned and $stride$ is
the number of consecutive units in that dimension when sharding. While existing systems implicitly assume that maximum
$stride$ is always used, equal to the size of the corresponding dimension divided by the number of shards.
However, we find that allowing for the use of other stride value may yield intuitive sharding strategies for certain models that contain
\texttt{Reshape} operations. For example, reshape is the meta description of tensor without any layout changing in memory, thus resharding communication which is inserted to connected the incompatible partitioning strategie with its input tensor is unnecessary. In our approach, we directly set the stride value of reshape with the value that the same as its input or output. Although the reshape may have an incompatible stride value, it has no side effects on the operation for the no influence on data layout. In this way, we can avoid introducing redundant resharding communications.

\subsection{Problem Definition\label{sec:ilp}}
 
We first analyze a simplified model sharding problem, which serves as the basis of \OurSystem's design.

Suppose we have a model with $n$ operators and the $i$-th operator has $s_i$ different sharding methods. The goal is to
find a sharding strategy $X$, where $X_{ij} = 1$ means we choose the $j$-th sharding method for the $i$-th operator.

We have two kinds of costs. The first is the computation cost $\text{comp}_{ij}$, the computation time of node $i$ when
using its $j$-th sharding method. The second kind of cost is the communication cost $\text{comm}_{i_1i_2j_1j_2}$ between
the operators $i_1$ and $i_2$ when there is link between $i_1$ and $i_2$ and they are using their $j_1$ and $j_2$
sharding methods, respectively.

The total cost can then be calculated as follows.

\vskip -1em
$$
C=\sum_{i=1}^n \sum_{j=1}^{s_i} \text{comp}_{ij}X_{ij} + \sum_{(i_1,i_2) \in E} \sum_{j_1=1}^{s_{i_1}} \sum_{j_2=1}^{s_{i_2}} \text{comm}_{i_1i_2j_1j_2}X_{i_1j_1}X_{i_2j_2}
$$

However, it contains a quadratic term $X_{i_1,j_1}X_{i_2,j_2}$ 
and cannot be directly solved with ILP solvers. Therefore,
we introduce auxiliary integer variables $A_{i_1i_2j_1j_2} \geq X_{i_1j_1} + X_{i_2j_2} - 1$ and rewrite the
optimization problem as follows ($[z]$ denotes the set $\{0, 1, \dots, z-1\}$). 

\noindent\begin{minipage}{\columnwidth}
    \small
    \begin{equation*}
        \min~~ \sum_{i=1}^n \sum_{j=1}^{s_i} \text{comp}_{ij}X_{ij} + \sum_{(i_1,i_2) \in E} \sum_{j_1=1}^{s_{i_1}} \sum_{j_2=1}^{s_{i_2}} \text{comm}_{i_1i_2j_1j_2}A_{i_1i_2j_1j_2}
    \end{equation*}
    \vspace{-3mm}
    \begin{align*}
        \text{s.t.:} &\quad \sum_{j=0}^{s_i} X_{ij} = 1, && \forall i \in [n] \\[-1mm]
                           &\quad A_{i_1i_2j_1j_2} \geq X_{i_1j_1} + X_{i_2j_2} - 1, && \forall (i_1,i_2) \in E, j_1 \in [s_{i_1}], j_2 \in [s_{i_2}] \\[-1mm]
                           &\quad X_{ij} \in \{0, 1\}, && \forall i \in [n], j \in [s_i] \\[-1mm]
                           &\quad A_{i_1i_2j_1j_2} \in \{0, 1\}, && \forall (i_1,i_2) \in E, j_1 \in [s_{i_1}], j_2 \in [s_{i_2}]
    \end{align*}
\end{minipage}


\subsection{Finding SPMD Strategies with Graph Coarsening}

\subsubsection{Finding Critical Nodes}
Modern DNNs may contain thousands of operations. Find-
ing the optimal solution to the ILP problem in Sec.~\ref{sec:ilp} at this
scale is often impractical. To reduce the problem size in the search algorithm, graph coarsening techniques are necessary. \OurSystem has developed effective graph coarsening heuristics inspired by data-driven feedback to aid online model training. After observing multiple online models, two important findings were ascertained: First, even though many operations are included, a few computationally-intensive nodes (e.g. MatMul and Convolution) account for the majority of the computing time. Second, the computation graphs of most models generally contains multiple parts that are computed concurrently. forms with a few nodes forming the backbone of the model. By identifying these computationally-intensive nodes on backbone and using them as segmentation boundaries, the entire graph can be divided into several subgraphs without impeding key solutions, while maintaining a certain degree of coarseness. Such computationally-intensive nodes on backbone are labeled as \textit{critical nodes}.

We define the backbone of the model as all operations on the critical path. To achieve this, we need to logically calculate each operation's earliest and latest calculation timings. Instead of calculating the real-time, we assume that the execution time of each operation is 1, and the tensor transmission time is 0. Then the earliest time for each operation can be calculated according to the topological order. On the contrary, the latest time of each operation is determined according to the reverse order of the topology. The backbones are the nodes with the same earliest and the latest time. Finally, we choose all the computationally-intensive nodes on the backbone as critical nodes and adopt them to coarsen the graph.

This method is developed based on a large amount of online feedback data and has been proven effective in practice, and there is still the possibility for improving this thumbnail strategy in the future due to data-driven feedback.

\subsubsection{Three-level Subgraph Merging\label{sec:subgraph_merging}}
We propose a novel method for deriving SPMD strategies by repeatedly merging subgraph. The core idea is divide and conquer: if we can
coarsen a computation graph into a single node, the optimal strategy can be easily determined by enumerating its
sharding methods. The challenge with this approach is to design a subgraph merging procedure that can
systematically reduce the graph size while preserving the majority of the strategy space. Based on the observations and
statistics from optimizing common models in a production cluster, we design a three-level subgraph merging algorithm for
searching SPMD strategies efficiently and near-optimally. We divide the entire graph by identifying special structures, namely
\textit{cone structures} and critical nodes, in DNN models, and merge these subgraphs with different
algorithms. Our methods is generalizable to new DNN models as we only recognize the graph structures rather than
specific operator types. On the other hand, our method utilizes the special structures that commonly appear in DNN models
and outperforms general heuristics that are oblivious to DNN training.

We first introduce the steps to merge a subgraph into a single node in \OurSystem. A
subgraph is represented as a set of nodes. \textit{Cut tensors} are those which connect the nodes in the subgraph with those outside the subgraph. Merging a subgraph into a single node requires us to remove the operators and tensors within the subgraph, insert a new abstract operator into the computation graph, and connect all cut tensors to the abstract operator. Further, the candidate strategies of the coarsen abstract operator is decided after subgraph merging, which relies on the merging algorithm.

\begin{figure}[t]
    \centering
    \includegraphics[width=0.48\textwidth]{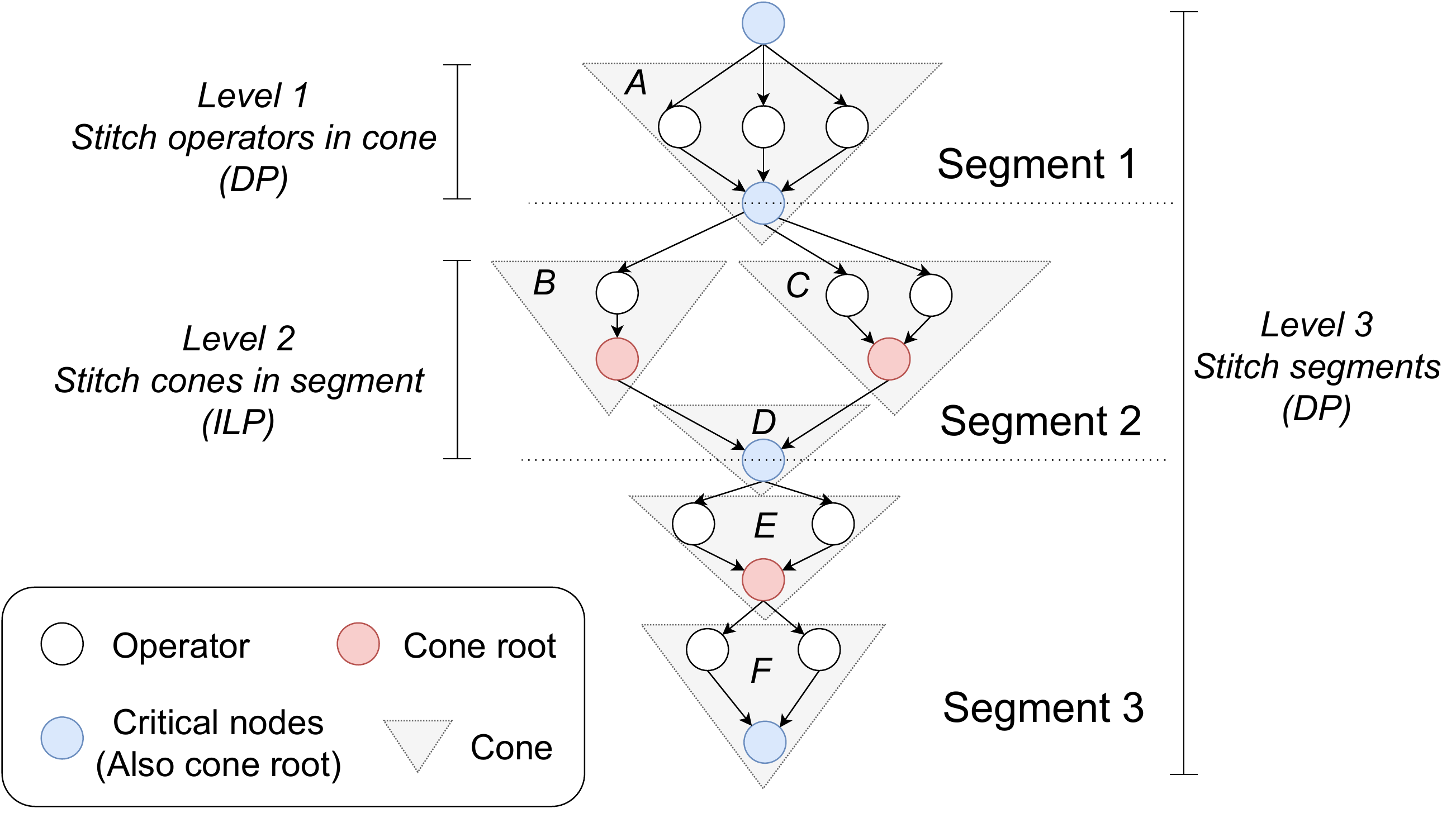}
    \caption{Hierarchical segmentation on graph and their matching merging methods.\label{fig:three_level}}
\end{figure}

We now introduce the three-level subgraph merging method, which is systematic designed for searching SPMD strategy. Macroscopically, \OurSystem divides model graph hierarchically into three levels. In each hierarchy, it determines SPMD strategies by employing either dynamic programming or ILP, as illustrated in the Fig.~\ref{fig:three_level}. For the bottom hierarchy (denoted as level 1), we should identify the \textit{cone structures} and \textit{cone root}. A cone root is an operator with multiple inputs. A cone structure is a special subgraph that comprises only one cone root and
other nodes. We observe that most DNN models in
our production cluster contain \textit{cone structures}. To identify cone structure, \OurSystem finds all cone roots in the entire graph, then extends each cone root with its inputs until reaching another cone root. Cone structures have the advantageous property that, inside a cone, all nodes other than the cone
root have only one input tensor. Otherwise, the node itself is a cone root. This facilitates the use of dynamic programming to stitch each operator's strategies by utilizing dynamic programming. Given the enumerated strategy of the cone root, \OurSystem optimizes the optimal strategy of all other nodes within the cone by minimizing communication cost introduced by resharding. Therefore, the strategy of the cone is determined due to the given strategy of cone root. We preserve all possible strategies of the cone for each enumeration in this hierarchy. In the middle hierarchy, the problem size 
has been significantly reduced since the graph comprises cone structures instead of fine-grained nodes. \OurSystem formulates ILP problem for merging cones with different strategies.

However, the number of cones can still be quite large, leading ILP solvers fails in acceptable time. This is the frequently encountered challenge for large models. \OurSystem introduces the top hierarchy and another heuristic to further reduce the problem complexity. It divides the entire graph into several near linear segments by using critical nodes as pivots, as illustrated in Fig.~\ref{fig:three_level}. As the result, the model graph is coarsened into a simple linear structure with possible strategies and can be easily stitched with dynamic programming. Within each segment, \OurSystem utilizes ILP solver to decide its strategy and employs dynamic programming to solve the cone strategy. It is the certain that strategy quality may be not guaranteed when introducing the top hierarchical segmentation. However, it is reasonable to reduce the exploration time when the strategy quality is not greatly affected. \OurSystem provides two optimization levels distinguished by whether introducing the top hierarchical segmentation, which contributes to the reduction of the problem complexity for cone merging. It provides flexible trade-offs between strategy quality and search time for different scenarios.


\subsubsection{Candidate Generation}

With our strided partitioning (Sec.~\ref{sec:sharding_spec}) approach, the strategy space is still too large to search
efficiently even with our three-level subgraph merging algorithm. For example, to put a 256x256 tensor onto 4 devices regularly,
there are $2 * \operatorname{factors}(256/4) + 2 = 16$ different possible methods, where the first $2$ refers to sharding along
the two dimensions, $\operatorname{factors}(256/4)$ is the number of possible strides when sharding on a dimension,
which is equal to the number of factors of the size on each device, and the last $2$ accounts for the non-sharding
methods, replication, and partial that the tensor has full shape but with partial result on each device.

An observation is that most of the sharding specs produce strategies of the same performance. For example, using
different strides throughout a model imparts strategies with the same amount of computation and communication.
Therefore, we propose that all tensors use the maximum strides by default, and only utilize other strides when they could
reduce communication. To this end, we design a candidate generation algorithm based on this idea.

Starting with a computationally-intensive operation (such as \texttt{MatMul}), we enumerate its sharding methods where
all involved tensors have maximum strides. Then we recursively merge it with a neighbor operation using the
comm-free merging algorithm (Sec.~\ref{sec:subgraph_merging}), until no neighbors can be merged without
introducing communication. During this process, all sharding methods used for each operator are recorded as candidates.
By limiting the search to these candidates, we can avoid exploring sharding specs that would not result in better strategies.


\subsection{Memory Constraints}

GPU memory limitation is a major concern when training large models with billions of parameters, and is one of the motivation
behind SPMD parallelism. By sharding parameter tensors and storing disjoint slices on different devices, the memory usage on
a GPU can be reduced. However, sharding typically introduces communication, thus lengthening the training time. Existing memory
saving techniques, such as ZeRO \cite{zero}, are applied on the whole graph. As a result, they often shard more tensors
than necessary, causing unnecessary communication overhead.

\OurSystem proposes an algorithm that automatically finds SPMD strategies that minimize the training time while ensuring
the memory usage is within the device limit. It treats memory usage as a constraint and do not over-optimize for it.
This can lead to complex strategies that are not easily designed manually. For example, it may apply ZeRO only on some
of the large tensors and replicating other tensors, when the memory limit is not enough to replicate all parameters.

Our algorithm is based on the observation that a small number of tensors occupy the majority of memory. Thus, sharding the large parameters can be utilized to save memory without causing a large impact on the generated strategy. Specifically, we first
sort the parameters in the model in descending order of their sizes and mark the first $k$ parameters. Then we remove
sharding candidates that replicate these tensors, forcing them to be partitioned during the strategy search. We choose
the smallest $k$ that provides a feasible solution.

\section{Pipeline Strategy\label{sec:pipeline}}
\subsection{Formulation\label{sec:pipeline_formulation}}
With SPMD parallelism, each device runs identical program on sharded tensors. When training on large clusters, the
sharded tensors on each device may become too small to saturate the computation units, causing GPU underutilization.
Pipeline parallelism can be adopted in complementary to SPMD parallelism to train very deep models.

Models are divided into stages when using pipeline parallelism. Each stage is then assigned to a group of devices and
further partitioned on the devices with SPMD parallelism. \OurSystem does not require sorting all instructions in linear order. Instead, it directly segments graph by utilizing the topology relationships for each node, thus the strategy quality is guaranteed.   \OurSystem, we consider pipeline parallelism on a dimension
of the device mesh to unify the strategy space with SPMD parallelism.

Suppose that we adopt pipeline parallelism on a device mesh dimension of size $d$. The first step is to divide the
model into $d$ stages. We solve the following optimization problem to find a stage assignment that minimizes the
communication volume while keeping the stages balanced.

The pipeline parallelism can be formulated as ILP problem. We use integer variable $B_i$ to represent the stage that operator $i$ is assigned to. For each edge $(i, j)$ in the
computation graph, we have $B_j - B_i \geq 0$, because the consumer of a tensor should be placed to a stage not earlier
than the producer of the tensor. When a producer-consumer pair is placed to different stages, i.e. $B_j \neq B_i$,
communication is required to transfer the tensor $(i, j)$ across devices. We introduce a 0-1 variable $F_{ij}$ for each
edge $(i, j) \in E$ to denote whether instructions $i$ and $j$ are placed to different stages. We use the big-M method
\cite{bigm} to define $F_{ij}$ by adding the following constraint:

$$
F_{ij} \leq B_j - B_i \leq F_{ij} \cdot M
$$

\noindent where $M \geq d - 1$ is a sufficiently large constant. When $i$ and $j$ are placed on the same stage, i.e. $B_i
= B_j$, $F_{ij}$ can only take 0 due to $F_{ij} \leq B_j - B_i$. When $i$ and $j$ are placed on different stages,
$F_{ij}$ must be 1 since $B_j - B_i > 0$ and $B_j - B_i \leq F_{ij} \cdot M$. $M \geq d - 1$ ensures that $B_j - B_i
\leq F_{ij} \cdot M$ always holds when $F_{ij} = 1$, as there are only $d$ stages and therefore $B_j - B_i \leq d - 1$.

With $F_{ij}$ defined, the objective of finding pipeline parallelism can be easily modeled to minimize the total communication volume as the following:

$$
\sum_{(i,j) \in E} F_{ij} \cdot \operatorname{tensor\_size}(i, j)
$$

\subsection{Bound Tightening}

\begin{figure}[t]
    \centering
    \includegraphics[width=0.3\textwidth]{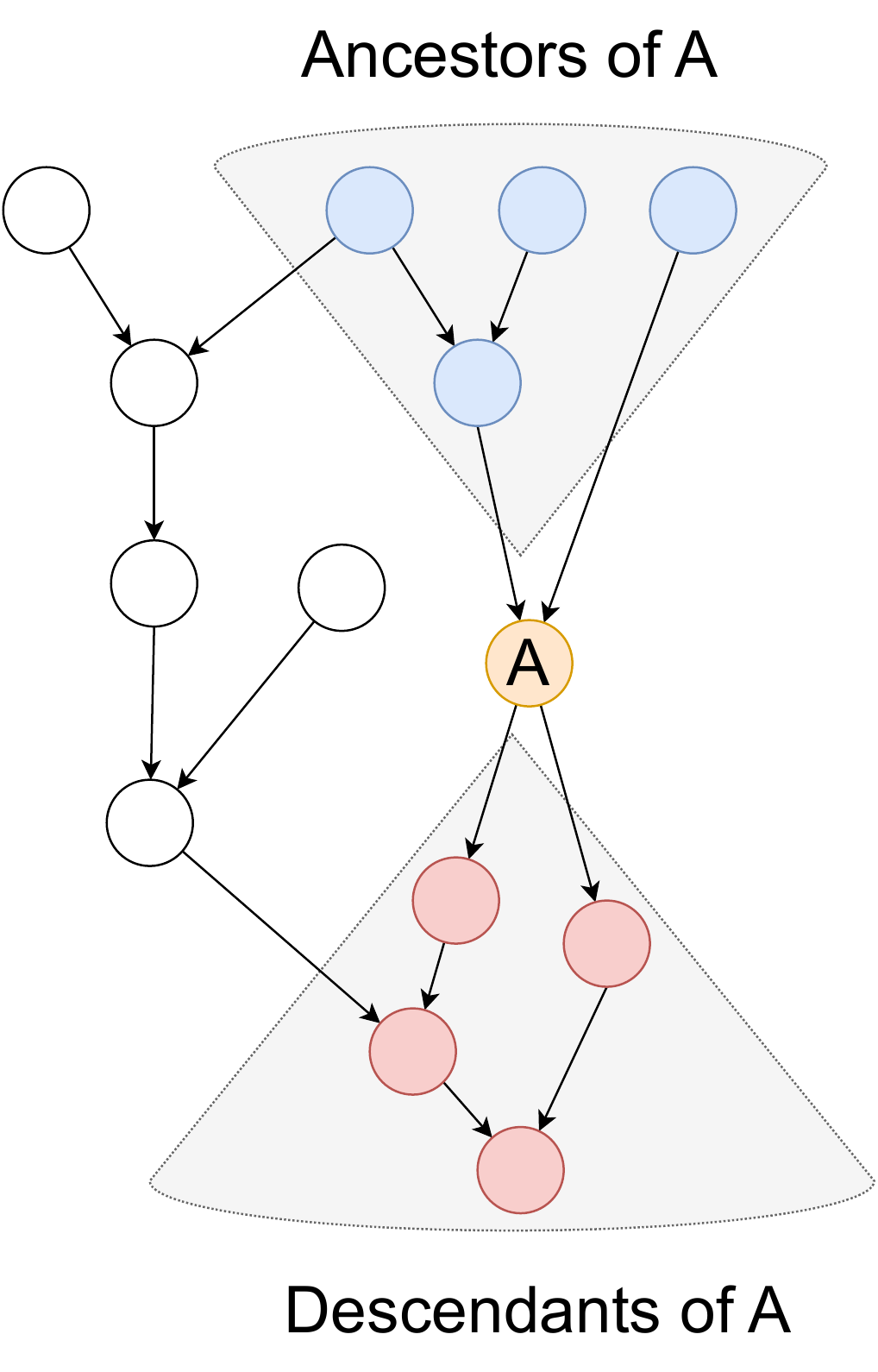}
    \caption{Ancestors and descendants of instruction A are represented in different colors. \OurSystem directly segments graph by topology relationships instead of sorting nodes in linear order.
 \label{fig:ancestors_descendants}}
\end{figure}

The formulation illustrated in Sec.~\ref{sec:pipeline_formulation} may have a daunting problem size for it comprises large amount of unbalanced stage divisions.
We propose a bound tightening technique to limit the maximal unbalanced tolerance for each stage to solve the optimization problem efficiently. For a given instruction
$i$ in the computation graph, we define two instruction sets $\text{A}_i$ and $\text{D}_i$  that are the ancestors and
descendants of $i$, respectively. An ancestor of $i$ is an instruction from which there exists a path to $i$ in the
computation graph. Similarly, an descendant of $i$ is an instruction whose ancestors includes $i$.
Fig.~\ref{fig:ancestors_descendants} illustrates the set $A_i$ and $D_i$.
The instructions in $\text{A}_i$ must complete before
executing $i$ due to data dependencies. Therefore, the accumulated execution time $\sum_{j \in A_i} comp_j$ is the
earliest start time of $i$. The universal instruction set is denoted as $M$. To maintain the balancing constraint, we have

$$
B_i \geq \lfloor \frac{\sum_{j \in A_i} comp_j}{time\_per\_stage} - \epsilon \rfloor
$$

\noindent where $\epsilon$ is a hyperparameter that specifies the tolerance level for imbalanced computation among
stages. $time\_per\_stage = \sum_{i \in M} comp_i / d$ is the computation time of each stage assuming perfect balance.
Conversely, we can also define the latest start time of $i$ based on $D_i$:

$$
B_i \leq \lceil \frac{comp_i + \sum_{j \in D_i} comp_j}{time\_per\_stage} + \epsilon \rceil
$$

To explain that, the two constraints limit the instruction $i$ to be placed on the stage which is previous or the successive of the perfect balance stage within the tolerance. After modeling these constraints in ILP formulation, the optimization problem can be solved efficiently. In \OurSystem, the tolerance level is provided by user to control the exploration time.

\section{Implementation\label{sec:implementation}}

\subsection{Task Graph\label{sec:task_graph}}

Existing systems such as Alpa\cite{alpa} adopt an MPMD-style runtime to orchestrate pipeline execution, which constructs
instructions for managing memory, synchronization, and utilizing a driver process to schedule instructions. In \OurSystem, we abstract Task Graph to arrange the strategy execution. A Task Graph is Directed Acyclic Graph (DAG) constituted of the following fundamental elements.

$\bullet$ \textbf{Task Node} A task node is used to represent the computations or Send/Recv p2p communication that take place across devices. When a pipeline is required, each task node has an associated stage HLO module, which can be regarded as a program while the task nodes represent its instances. For applications requiring Multiple SPMD-style tasks share the same HLO module is offseted with various inputs or parameters.

$\bullet$ \textbf{Source and Sink node}. They represent the start and the end node of Task Graph.

$\bullet$ \textbf{Edge}. It represents a series of tensors transmission between task nodes.

\begin{figure*}[t]
    \centering
    \includegraphics[width=.9\textwidth]{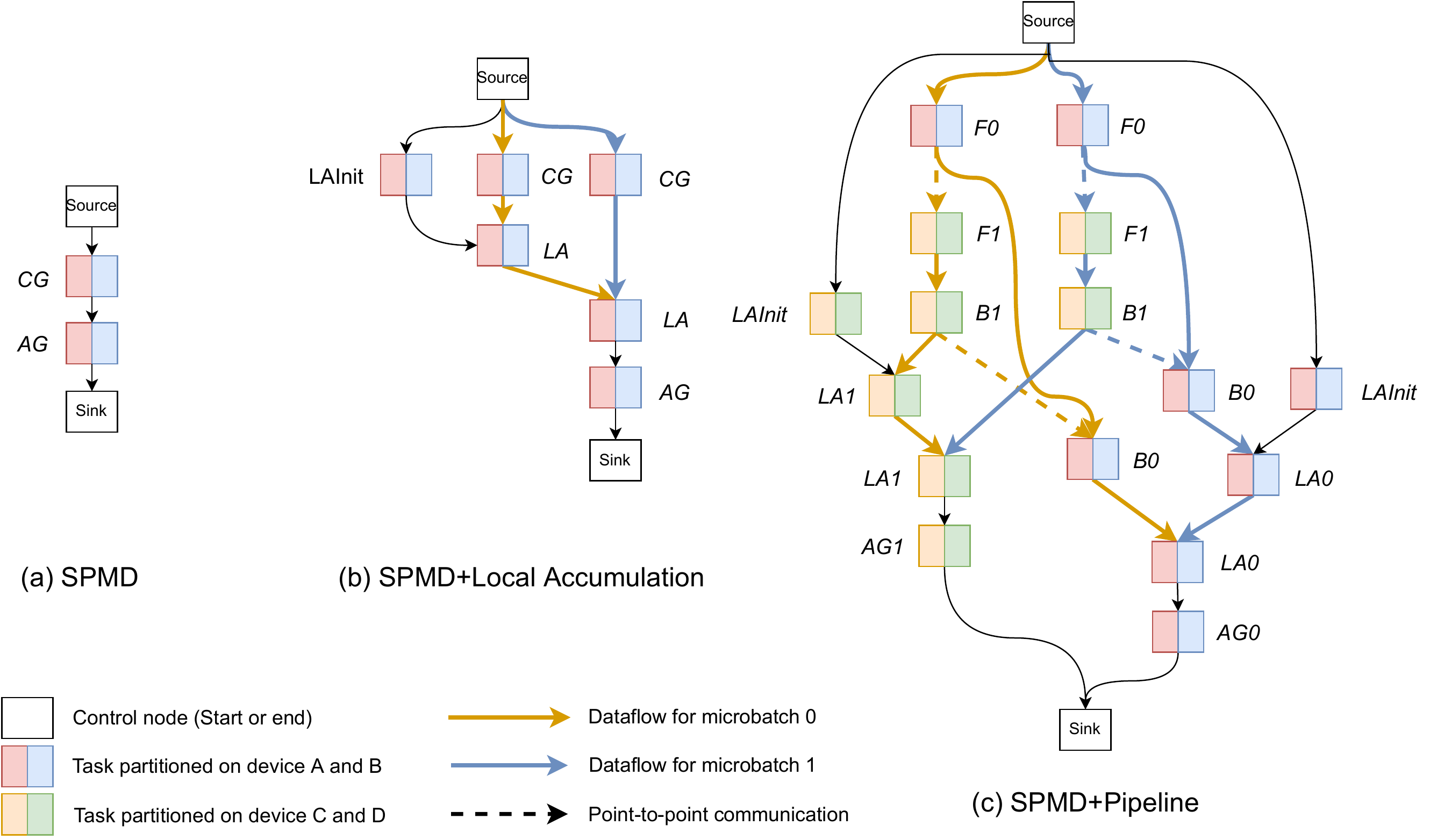}
    \caption{Task Graph with different strategies. Different colors represent different devices. Nodes with multiple colors are sharded across multiple devices. Blue and Yellow lines means different micro batches respectively. Dashed lines mean cross devices p2p communication where Send/Recv pair should be inserted while solid lines are not. For pure SPMD strategy (a), task graph only contains compute gradients(CG) and apply gradients(AG). For SPMD combined with multiple micro batches (b), the local accumulation(LA) should be created. For SPMD combined with Pipeline Strategy, the local accumulation should also be created with their corresponding backward stages. \label{fig:task_graph}}
\end{figure*}

Task graph provides an intuitive way to organize and execute computations or communications between task nodes. Generally, such computation is described in a DAG, which consists of multiple task nodes. Multiple task nodes can share the same HLO module, feeding with different inputs or parameters to enable SPMD parallelism. It can also implement pipeline parallelism by connecting multiple task nodes in a sequence. With these task graphs, memory saving scheduling and gradient checkpointing can also be planned automatically. Task Graph with different strategies is displayed in Fig.~\ref{fig:task_graph}.

\subsection{Static Scheduling\label{sec:static_scheduling}}

Static scheduling is a form of parallel computing which involves assigning pre-defined tasks to different processing elements. In comparison to dynamic scheduling, which involves dynamically assigning tasks while they are running, static scheduling offers better performance since the order and tasks are already pre-defined and do not need to rely on a master to decide what to do next. Currently, the 1F1B\cite{pipedream,dapple} schedule order is
implemented in \OurSystem as the default plan, which is proven necessary to improve throughput while saving memory. Specifically, task graph-based static scheduling is easier to implement and has the potential to explore better plans. This is because task graph is coarse-grained, resulting in a much smaller scheduling searching space than HLO graph, thereby reducing the time needed for exploration.

\subsection{Distributed Initialization and communicators management}
The variables initialization is very time-consuming when training large models. To facilitate this procedure, \OurSystem implements distributed initialization directly on assigned devices, rather than initializing on the master node and copying relevant shards to the destination. Ensuring numerical correctness during initialization relies on skipping to the correct seed state, based on the sharding information of the tensor.

\OurSystem's runtime is able to run multiple parallel strategies simultaneously, making the management of communicators highly complex. To simplify this process, we create and negotiate all required communicators for the current training model before the training begins. We gather the communication instructions, along with Send and Recv task nodes, and attempt to create the appropriate communicators. To prevent duplication, a global communicator cache are maintained.




\section{Evaluation\label{sec:evaluation}}


\subsection{Experiment Setup}


\textbf{Testbed}. We run our experiments on two clusters with typical commodity hardware.

\begin{itemize}
    \item \textbf{Platform M8:} (each machine) 8 * V100-SMX2-32GB GPU w/NVLink2, 2 * Xeon (Skylake) CPU 48C 2.5GHz, 768GB
    DDR4-2666, 1 * 100G RoCE interconnect.
    \item \textbf{Platform S1:} (each machine) 1 * V100S-PCIE-32GB GPU, 2 * Xeon (Cascade Lake) 52C 2.5GHz, 512GB
    DDR4-2666, 1 * 100G RoCE interconnect.
\end{itemize}

All machines are installed with Linux kernel version 4.19.91, Nvidia driver 470.82 and NCCL library 2.8. For different
baseline systems, CUDA toolkit with different versions (Alpa/DeepSpeed: 11.4, \OurSystem: 10.1) are used due to different
software requirements.

\textbf{Benchmark models}. We mainly experiment with two NLP models, GPT\cite{gpt3} and GShard MoE\cite{moe}, when
comparing to baselines due to limited open-source implementation of other models on the baseline systems. The
configurations used for these models are listed in Table \ref{tab:gpt_conf} and Table \ref{tab:moe_conf}. We
choose the configurations to be the same as those used by related studies. We also evaluate \OurSystem with VGG19
\cite{vgg}, DNABERT \cite{dnabert}, UNet \cite{unet}, and Wide-ResNet \cite{wideresnet}.

\textbf{Baseline systems}. We compare \OurSystem with two state-of-the-art systems, Alpa \cite{alpa} and
Megatron/DeepSpeed \cite{megatron,deepspeed}. Alpa is a automatic distributed training system that supports SPMD
parallelism and pipeline parallelism. Megatron/DeepSpeed means running Megatron's expert-designed parallelization
strategy and optimized Transformer implementation on DeepSpeed.

All experiments are performed with full FP32 (single-presicion) computation.

\begin{table}
\centering
\caption{GPT Model weak scaling test configurations. Global batch size is set to 64 for full parallelism tests, while 4 for SPMD-only tests. "GPT-3 Medium" configuration is used for SPMD-only 1-8 card strong scaling test.\label{tab:gpt_conf}}
\resizebox{\columnwidth}{!}{
\begin{tabular}{lccccccc}
\toprule
Config Name & $N_{params}$ & $N_{layers}$ & $D_{hidden} $ & $D_{ffn}$ & $N_{heads}$ & $D_{head}$ & $N_{GPUs}$ \\
\midrule
GPT-Medium & 350M & 24 & 1024 & 4096 & 16 & 64 & 1 \\
GPT-Large & 760M & 24 & 1536 & 6144 & 16 & 96 & 2 \\
GPT-XL & 1.3B & 24 & 2048 & 8192 & 32 & 64 & 4 \\
GPT-2.7B & 2.7B & 32 & 2560 & 10240 & 32 & 80 & 8 \\
GPT-6.7B & 6.7B & 32 & 4096 & 16384 & 32 & 128 & 16 \\
\bottomrule
\end{tabular}}
\end{table}

\begin{table}
\centering
\caption{GShard MoE Model weak scaling test configurations. Global batch size is set to 256 for full parallelism tests, while 8 for SPMD-only tests.\label{tab:moe_conf}}
\resizebox{\columnwidth}{!}{
\begin{tabular}{lcccccccc}
\toprule
Config Name & $N_{params}$ & $N_{layers}$ & $D_{hidden} $ & $D_{ffn}$ & $N_{heads}$ & $D_{head}$ & $N_{expers}$ & $N_{GPUs}$ \\
\midrule
MoE-8E-380M & 350M & 8 & 768 & 6144 & 16 & 48 & 8 & 1 \\
MoE-8E-760M & 760M & 16 & 768 & 6144 & 16 & 48 & 8 & 2 \\
MoE-16E-1.3B & 1.3B & 16 & 768 & 6144 & 16 & 48 & 16 & 4 \\
MoE-16E-2.4B & 2.4B & 16 & 1024 & 8192 & 16 & 64 & 16 & 8 \\
MoE-32E-10B & 10B & 16 & 1536 & 12288 & 16 & 96 & 32 & 16 \\
\bottomrule
\end{tabular}}
\end{table}

\subsection{Scalability}

\begin{figure*}[t]
	\centering
	\subfloat[] {
		\label{fig:gpt_m8}
		\includegraphics[width=0.4966\textwidth]{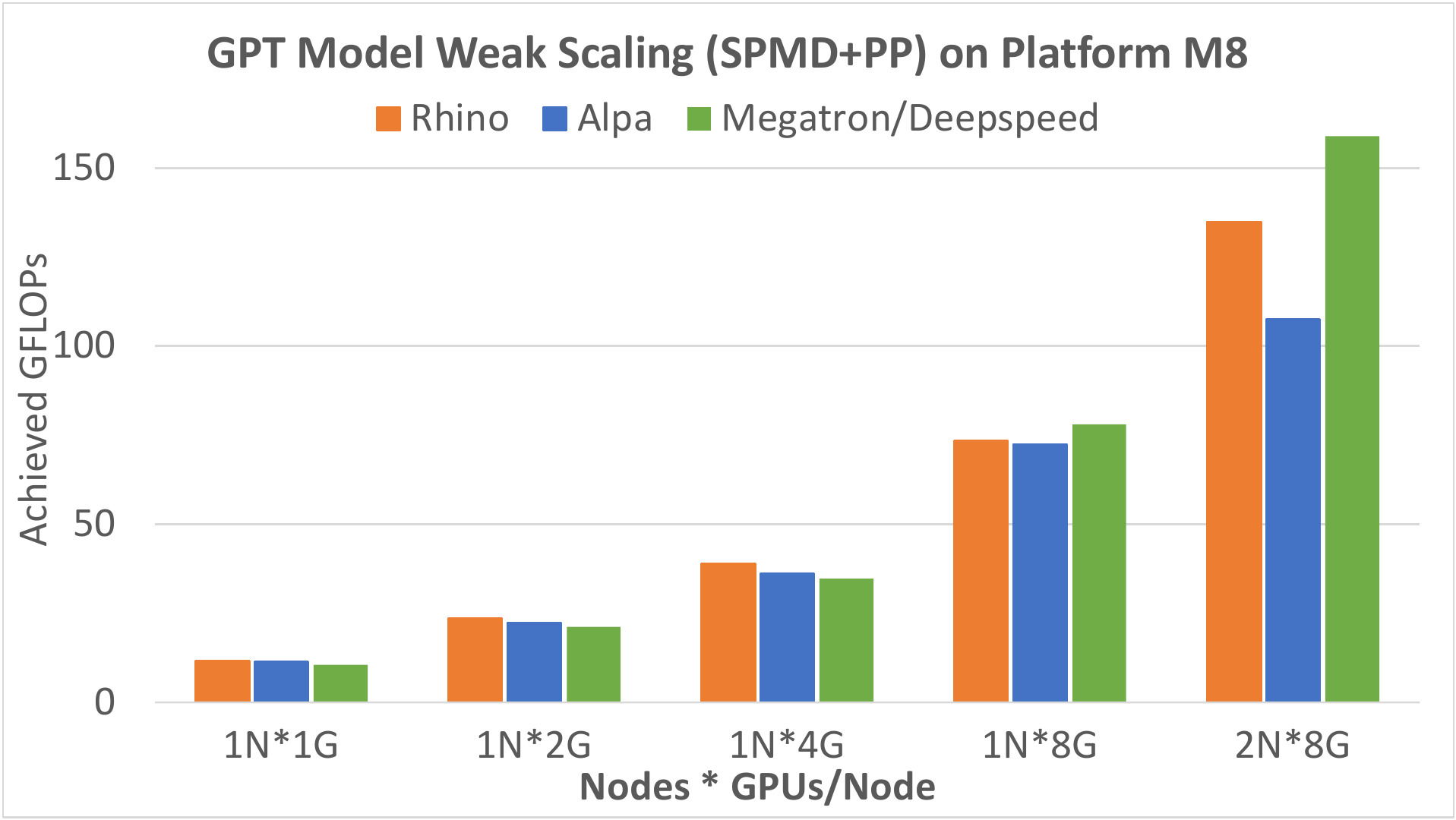}}
	\subfloat[] {
		\label{fig:moe_m8}
		\includegraphics[width=0.483\textwidth]{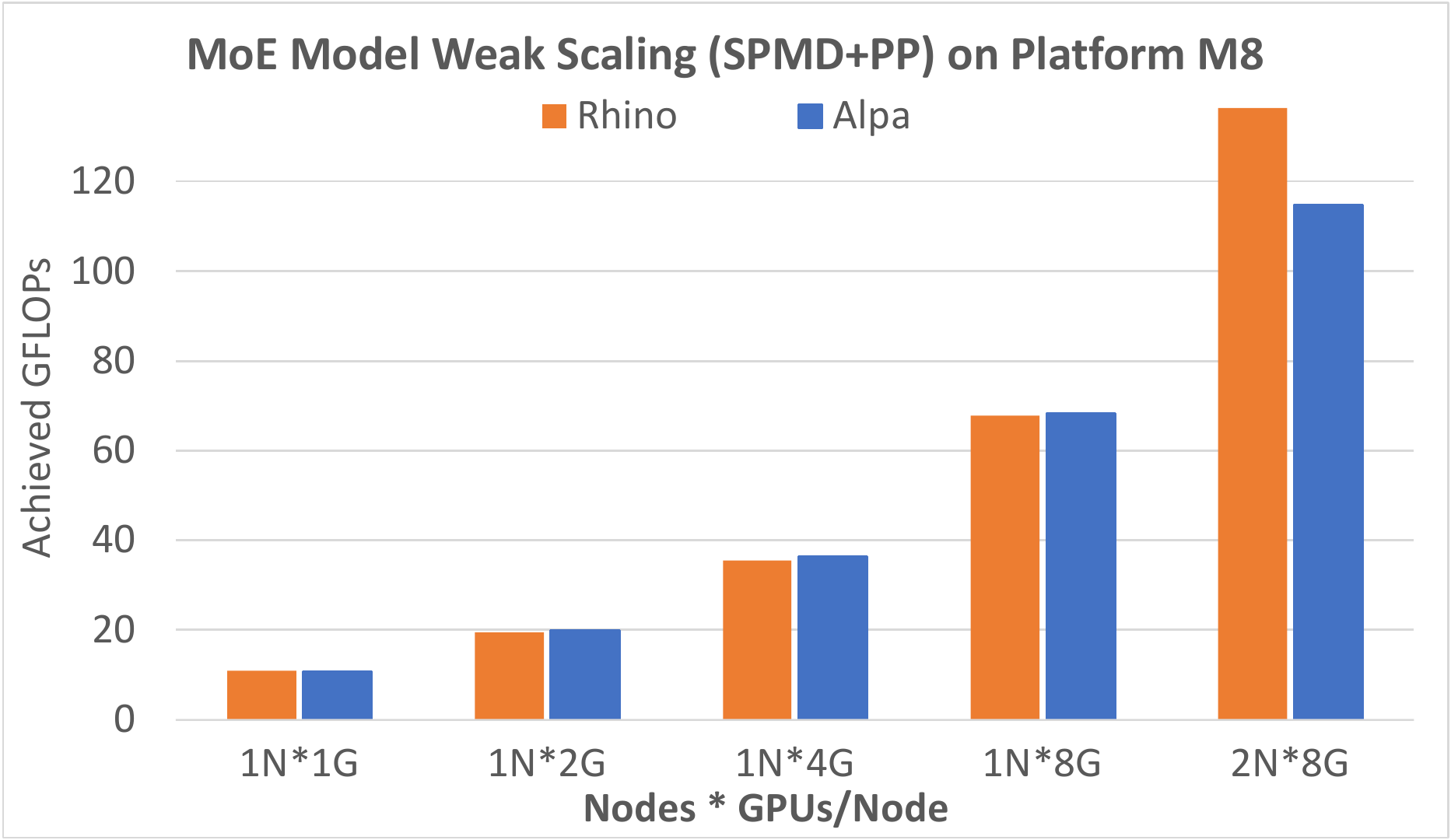}}
	\caption{GPT and MoE model scaling (by arguments) performance on Platform M8, with full parallelism.\label{fig:spmd_pp_m8}}
\end{figure*}

We evaluate the scalability of \OurSystem and baselines with up to 16 GPUs on Platform M8 for the GPT and MoE models. The
results are shown in Fig. \ref{fig:spmd_pp_m8}.

\begin{figure*}[t]
	\centering
	\subfloat[] {
		\label{fig:gpt_spmd_m8}
		\includegraphics[width=0.39\textwidth]{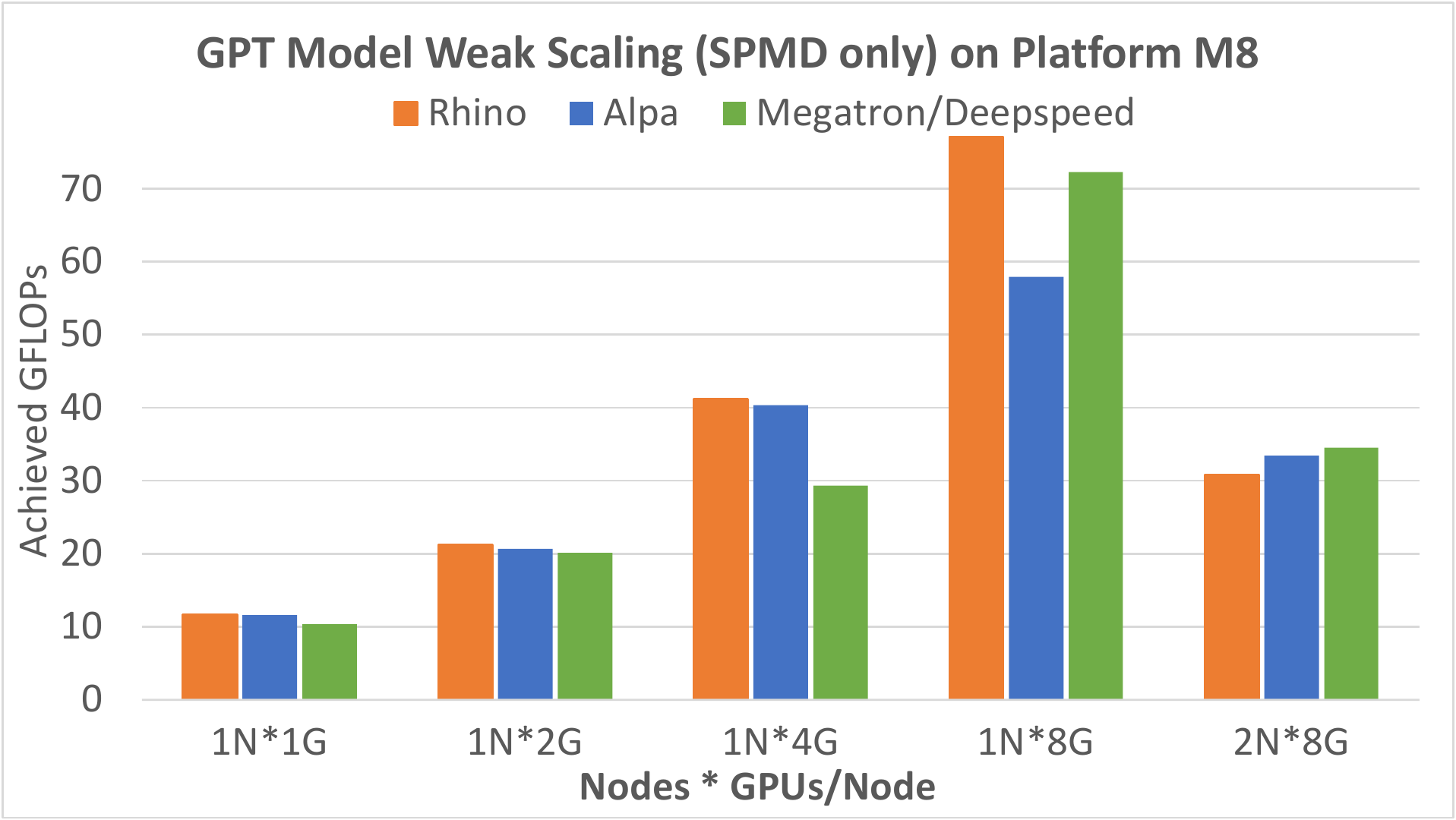}}
        \qquad
	\subfloat[] {
		\label{fig:moe_spmd_m8}   
		\includegraphics[width=0.39\textwidth]{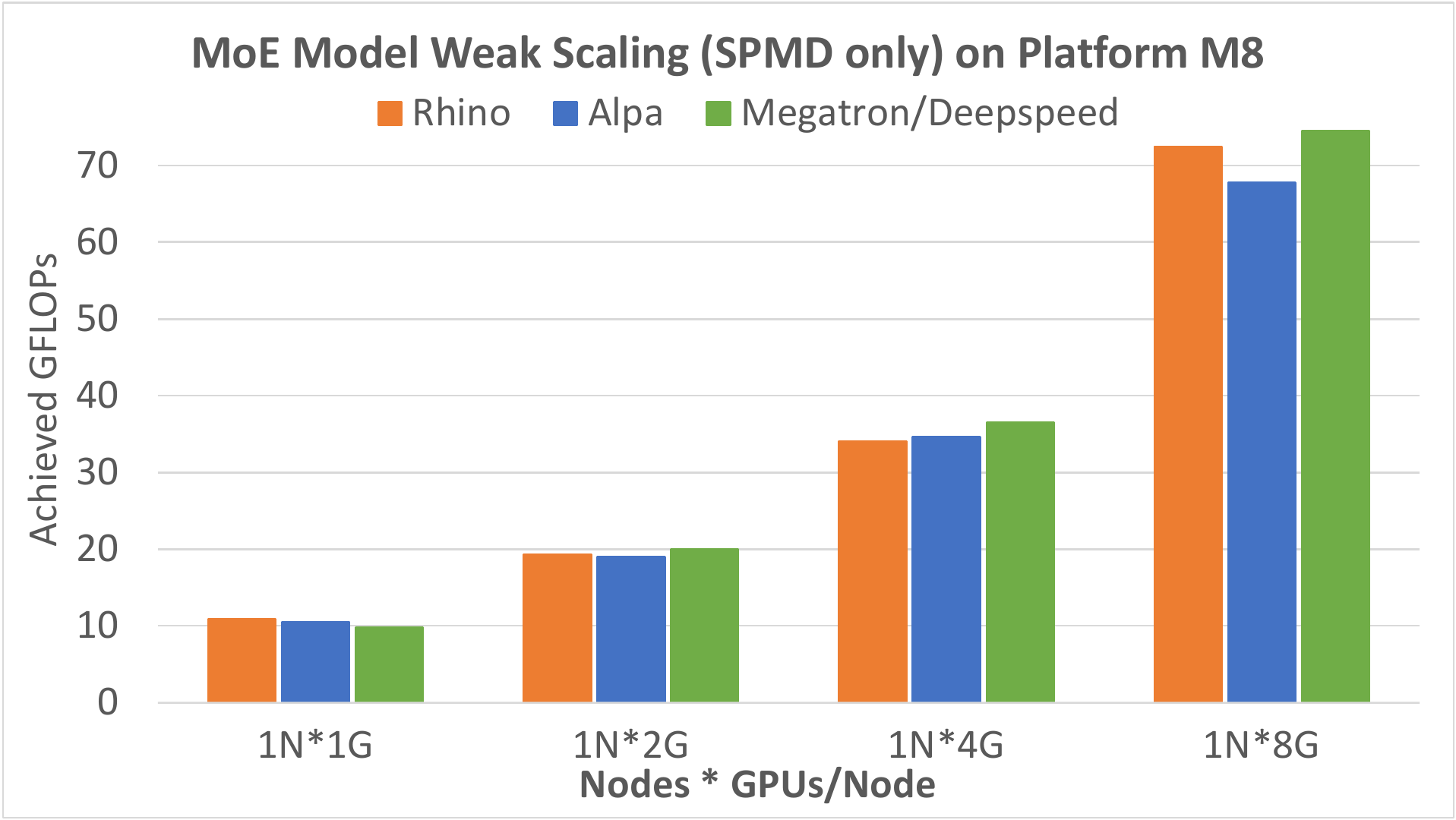}}

	\subfloat[] {
		\label{fig:gpt_spmd_s1}    
		\includegraphics[width=0.259\textwidth]{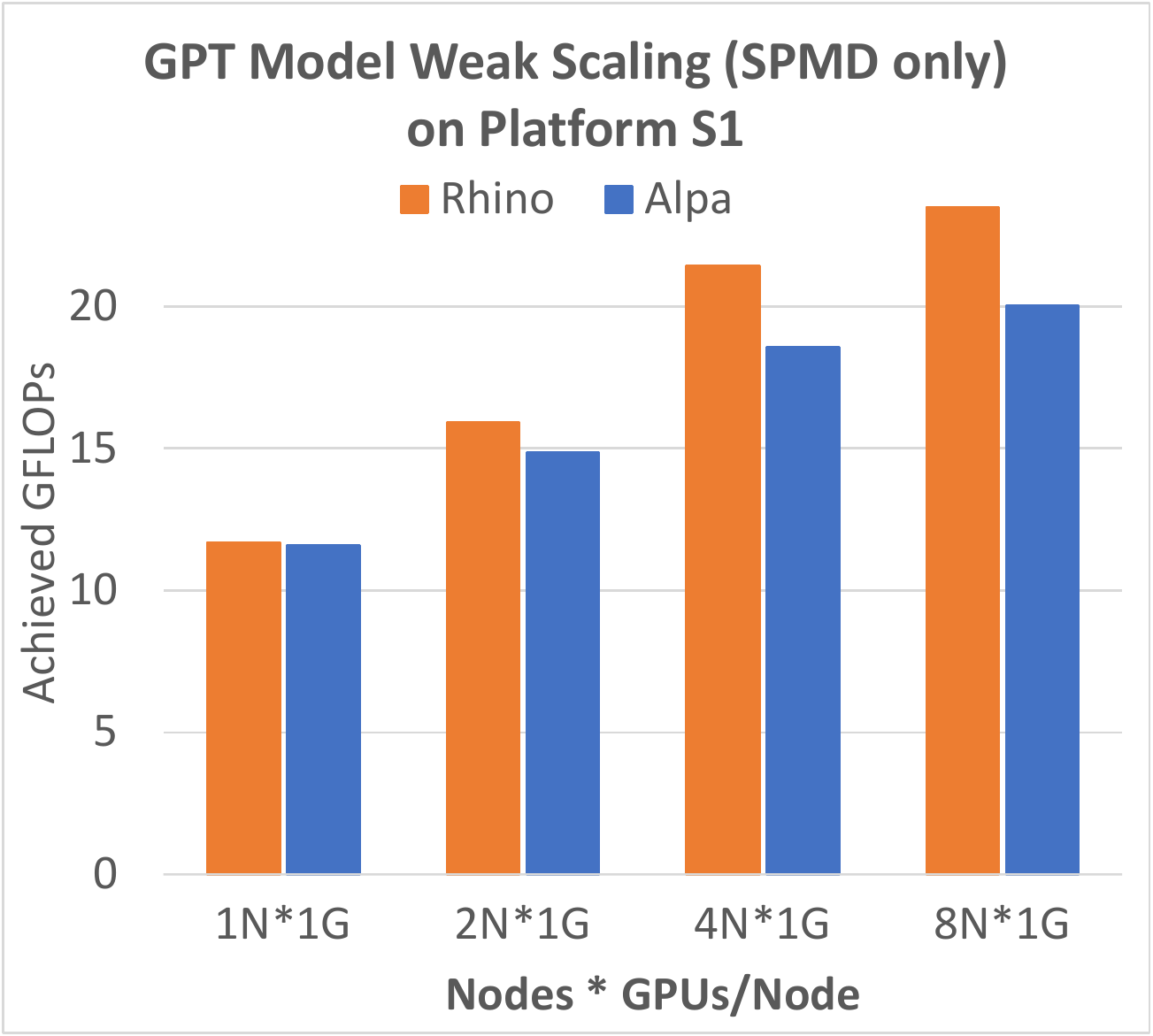}}
	\subfloat[] {
		\label{fig:gpt_pp_s1}    
		\includegraphics[width=0.259\textwidth]{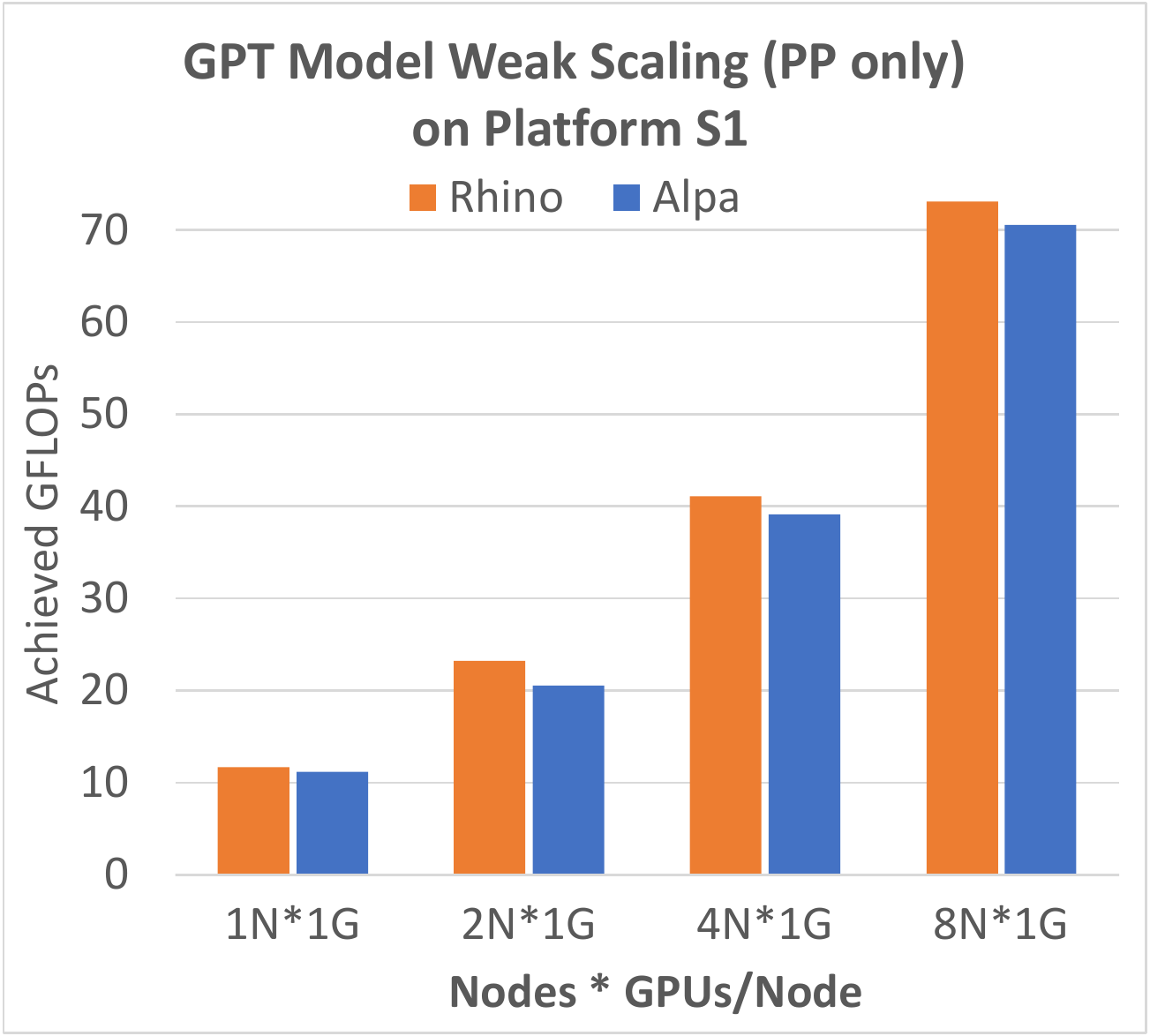}}
	\subfloat[] {
		\label{fig:gpt_strong}
		\includegraphics[width=0.397\textwidth]{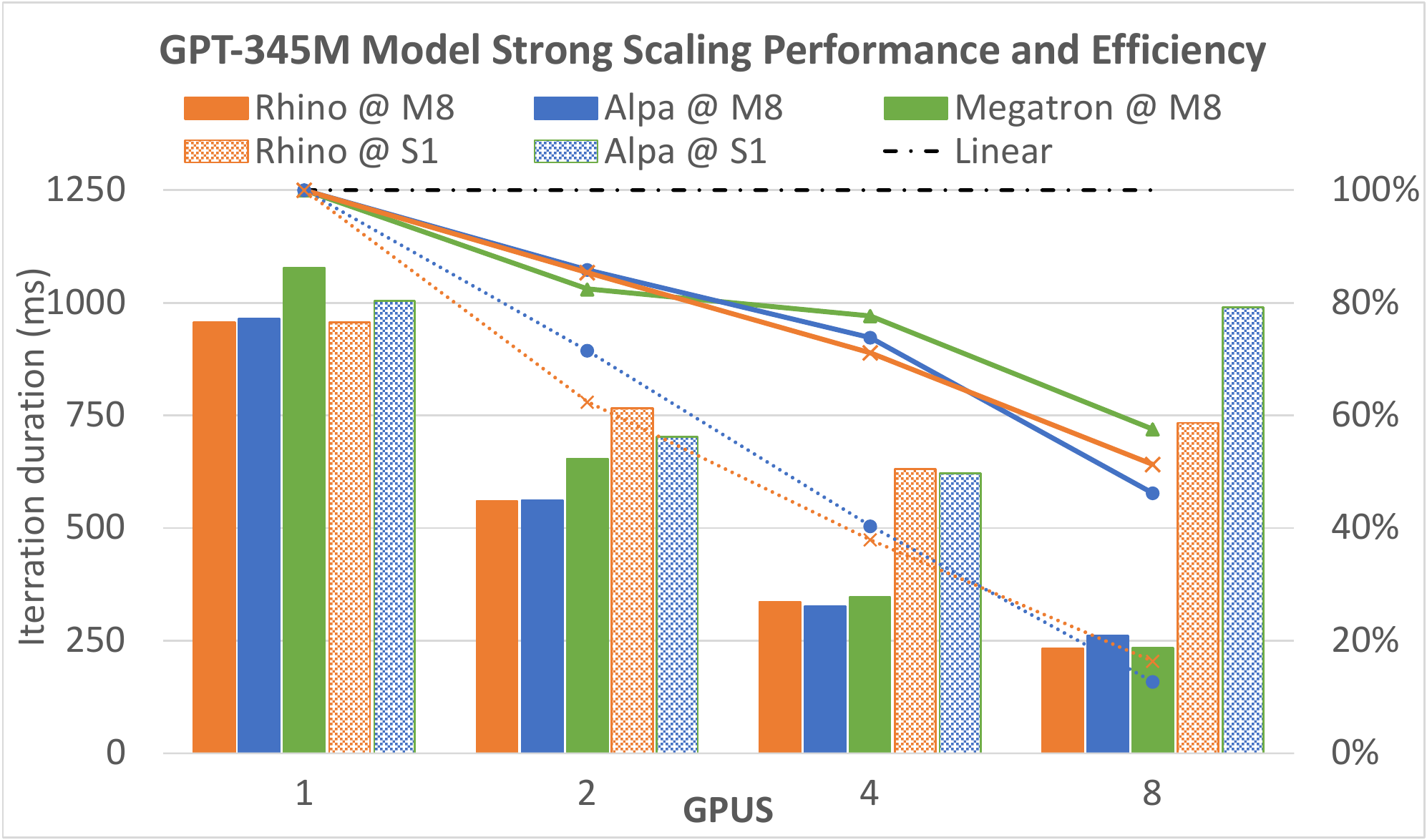}}
	\caption{GPT amd MoE model scaling performance with SPMD or PP parallelism only. (a)-(d) are arguments weak scaling cases mentioned in Table \ref{tab:gpt_conf}, (e) is strong scaling on GPT-Medium configuration. \label{fig:spmd_only}}
\end{figure*}

\begin{figure*}[!ht]
    \centering
    \subfloat[VGG-19] {\includegraphics[width=0.5\textwidth,height=15em]{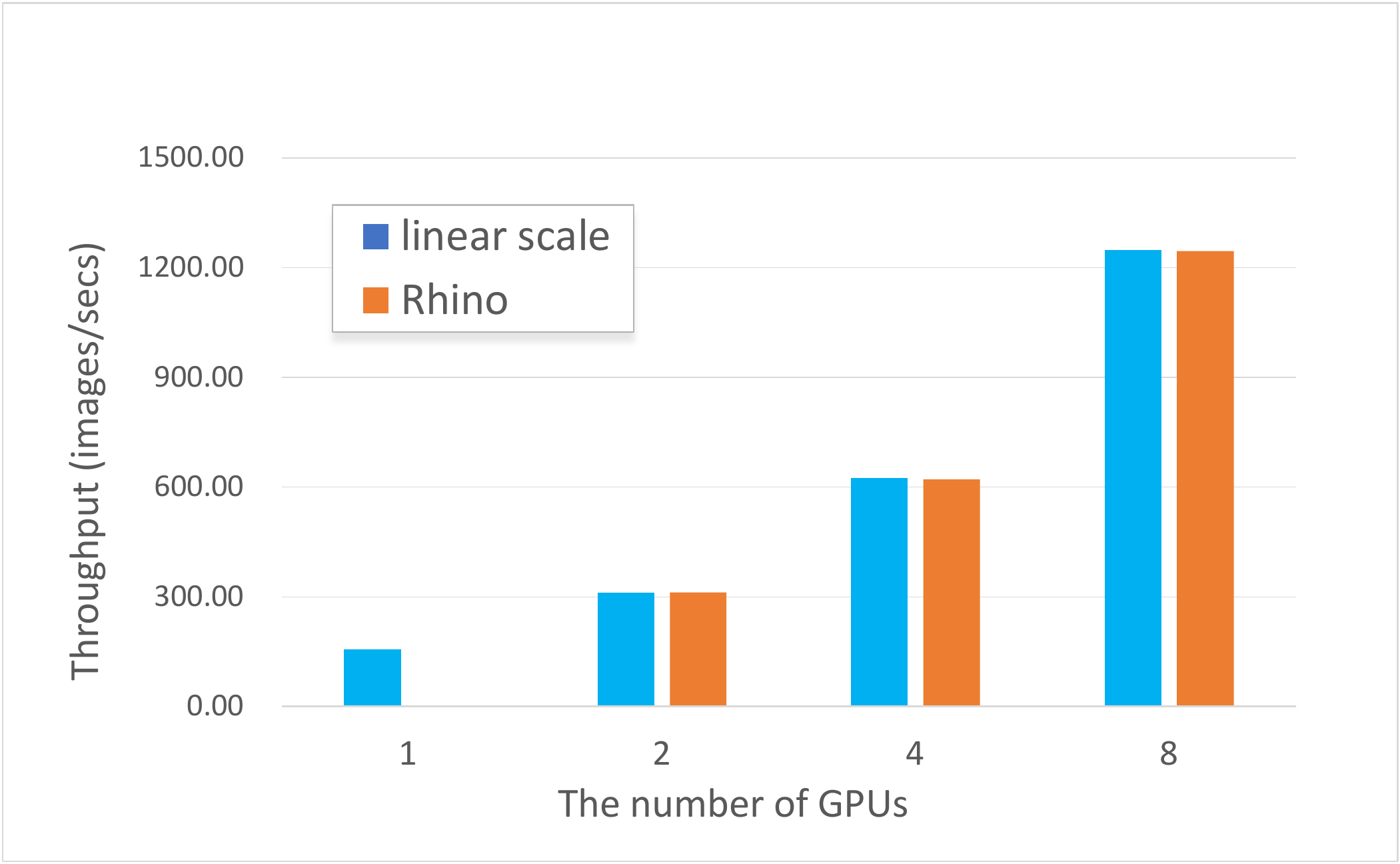}}
    \subfloat[DNABert]{\includegraphics[width=0.5\textwidth,height=15em]{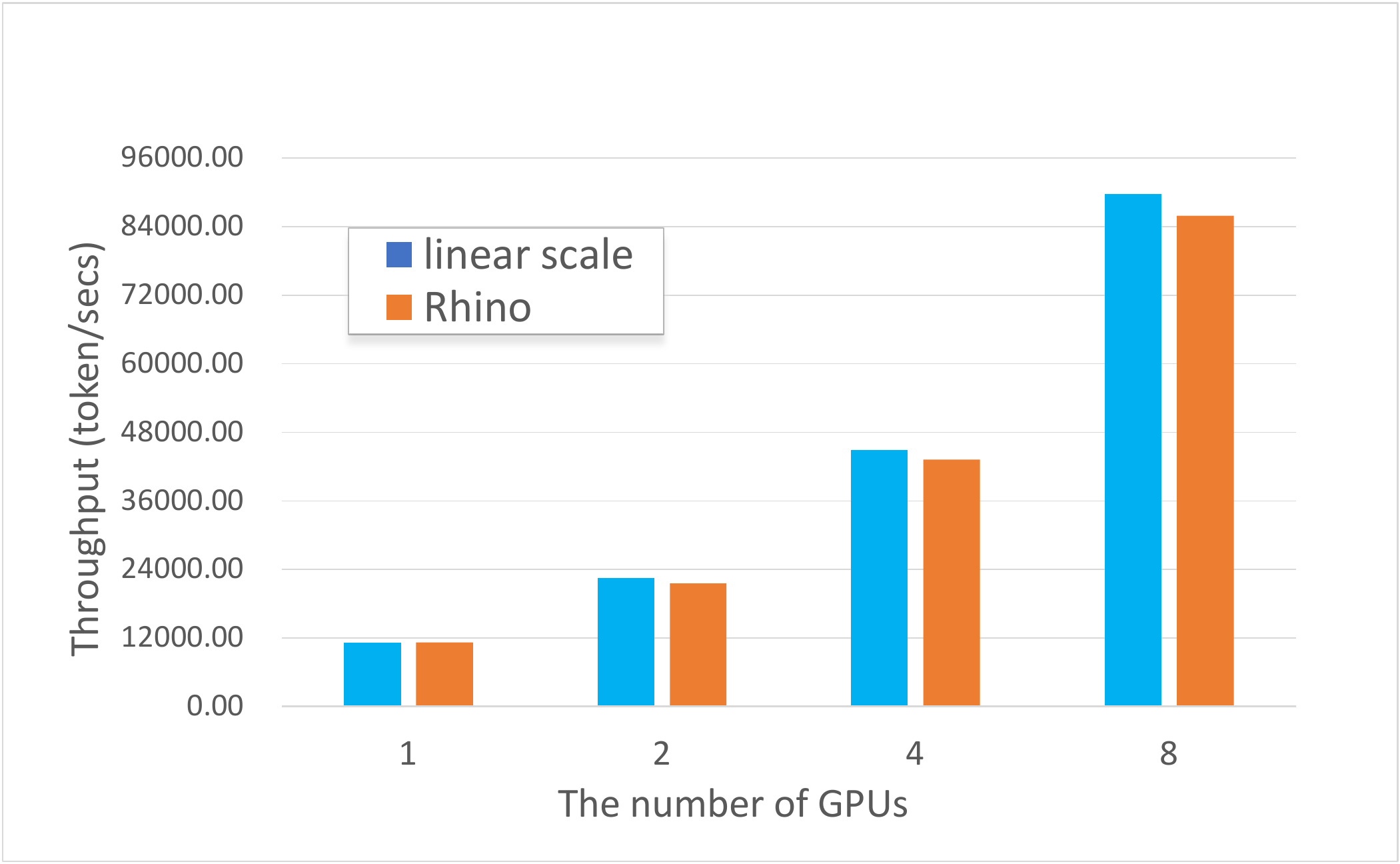}}
    \newline
    \subfloat[UNet]{\includegraphics[width=0.5\textwidth,height=15em]{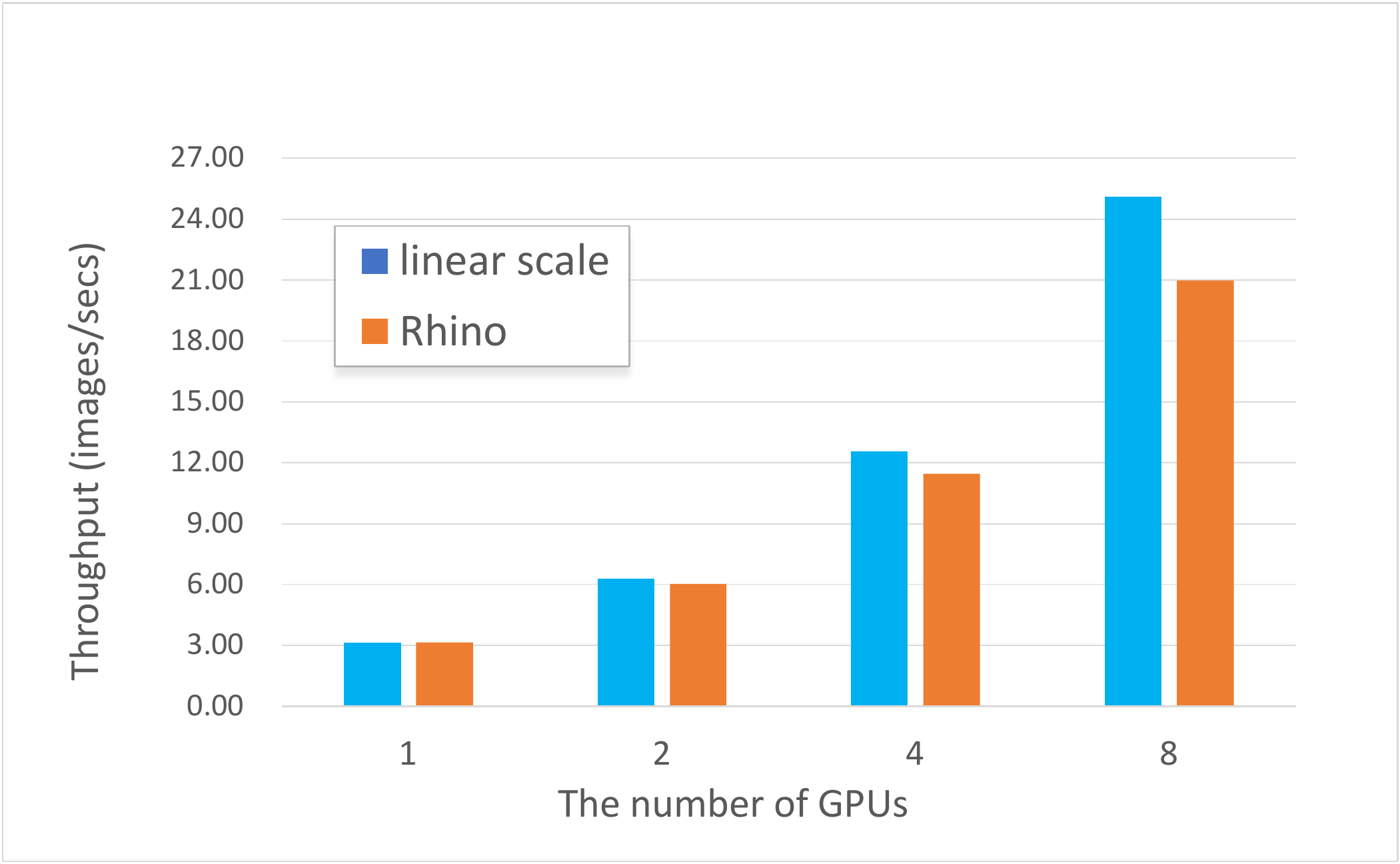}}
    \subfloat[Wide-ResNet]{\includegraphics[width=0.5\textwidth,height=15em]{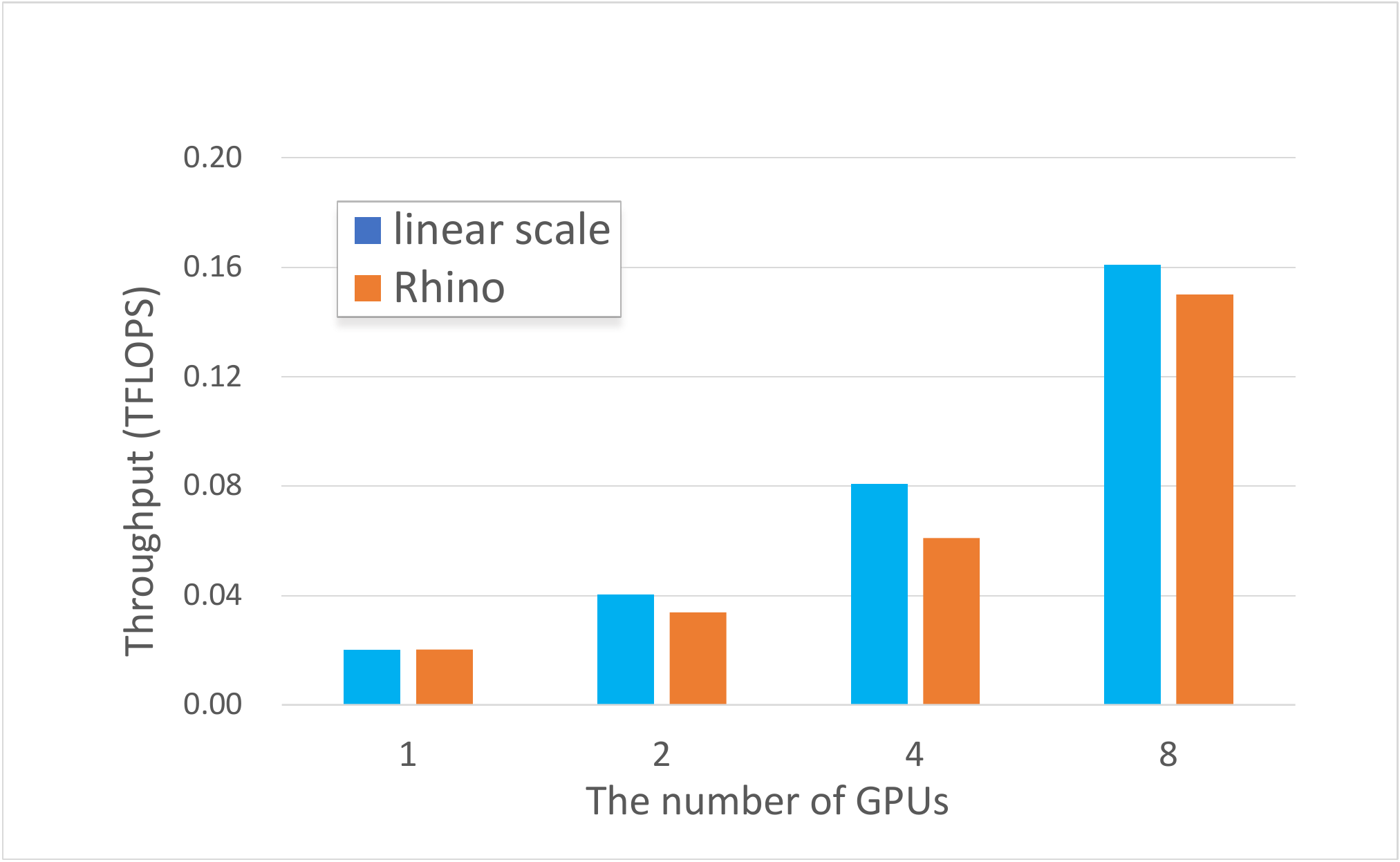}}
    \caption{
    Weak scaling evaluation of other general models on single node of M8.
    }
    \label{fig:show_cases}
\end{figure*}

\begin{figure}[t]
    \centering
    \subfloat[]{
    \label{fig:opt_time}
    \includegraphics[width=0.5\textwidth]{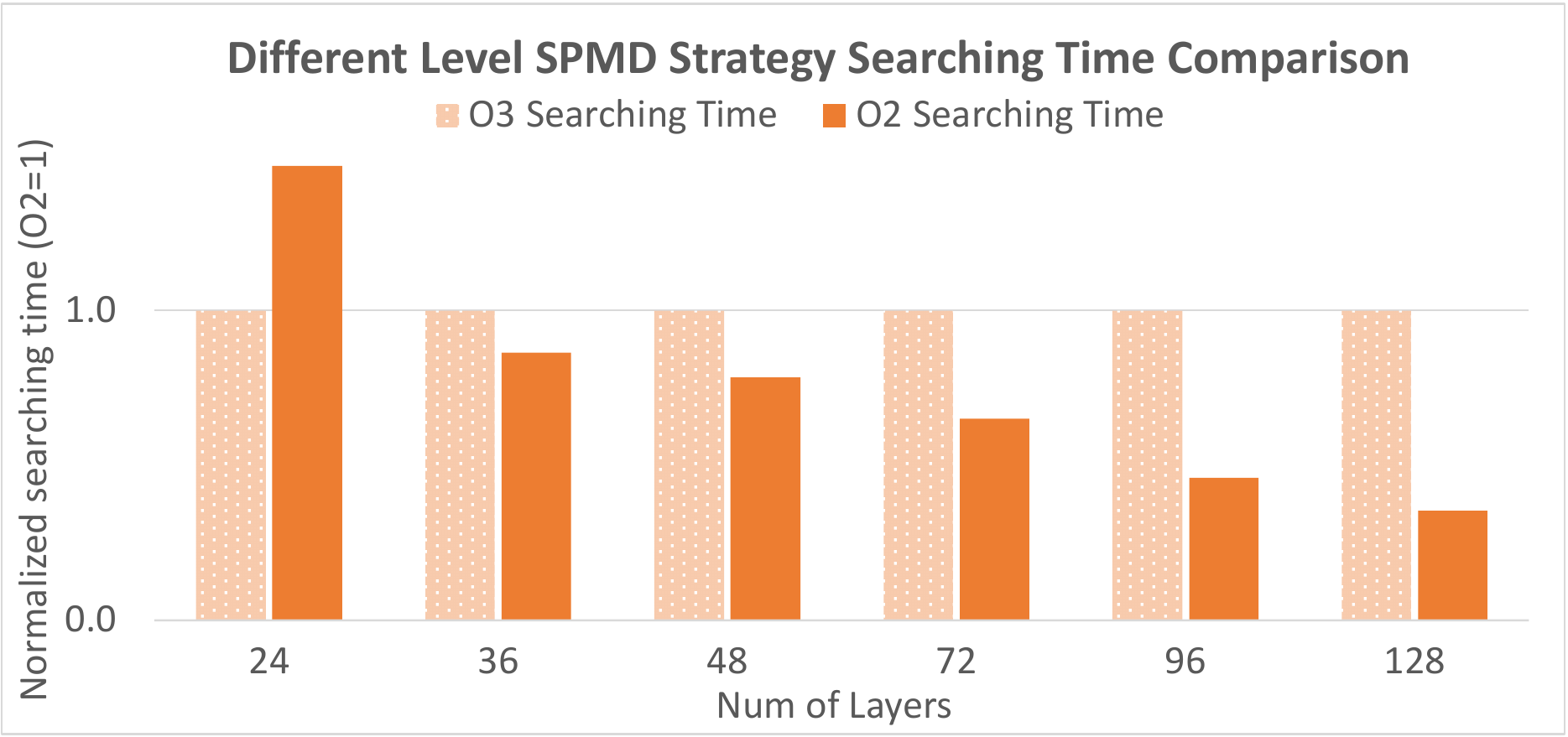}}
    \newline
    \subfloat[]{
    \label{fig:opt_perf}
    \includegraphics[width=0.5\textwidth]{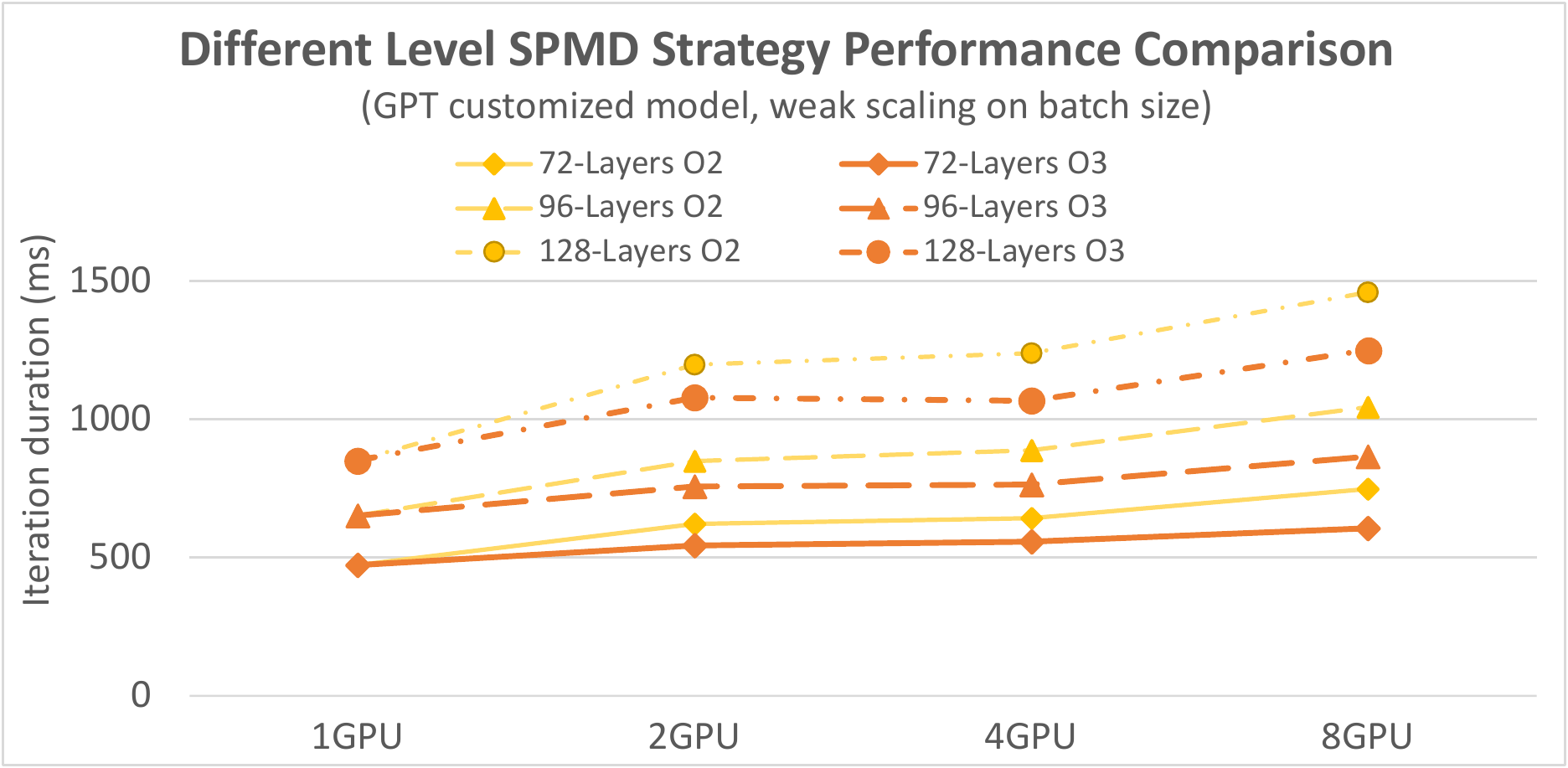}}
    \caption{Comparison for different level SPMD strategy searching. 'O3' denotes to full graph ILP method, while 'O2' means subgraph ILP plus DP method.\label{fig:opt_level}}
\end{figure}

The \textbf{\textit{\uppercase{fully automatic}}} parallelism is the essential desgin criterian of \OurSystem, that
enables exploring of the most efficient strategies of data, tensor (or sharing) and pipeline parallel mixture. The
system ability is validated by scaling tests from 1 to 16 GPUs (2 nodes on M8) with the fully parallelism strategy
planning for GPT and GShard MoE, results shown in Fig. \ref{fig:spmd_pp_m8}. Same workloads were also run on the same
platform with Alpa and DeepSpeed, results shown in the figure. For Alpa, we use 'PipeShardParallel' and try to find the
benchmark settings in its given examples. While for DeepSpeed, we tested some feasible parallel scheme settings and
record the best one.

It could be observed from the results that GPT and MoE model training driven by \OurSystem scaled well both inside a
single node (1 to 8 GPUs) or across two nodes. We measured the real floating-point arithmetic operations reached for
each test along the method of \cite{megatron}. Considering the ideal peak single-precision performance of V100-SMX GPU,
15.6 TFLOPs, the MFU \cite{reducear} of GPT model is 54.0\% to 75.9\% \OurSystem training, for MoE model it's upto
55.1\%. Compared with tests driven by another fully auto-parallel work Alpa and the state-of-the-art manually parallel
LM training framework Megatron (adopted within Deepspeed), \OurSystem provides equivalent performance with Alpa and
DeepSpeed/Megatron in many cases, sometimes slightly better. 

\subsection{Ablation study}
A significant portion of parallel training tasks use less than 8 GPUs, which could be observed in the statistics of our real production platform. Although we designed the system architecture for SPMD+PP parallelism, its performance on SPMD-only cases could be also be examined for those general popular models and workloads.

We performed weak scaling SPMD-only tests for GPT and MoE models, along the configurations in Table \ref{tab:gpt_conf} and Table \ref{tab:moe_conf}. Furthermore, strong scaling for GPT-Medium config was also done, in order to get better understand for SPMD strategy auto planning results. Same tests were also run with Alpa and Megatron/DeepSpeed systems. For Alpa, the default 'ShardParallel' method was used in SPMD-only test case. Manually tuned DP and MP parallelism arguments were carefully set to get the best result under Megatron/DeepSpeed cases. Most of the tests were done both on Platform M8 and S1. Considering there are a lot of scheduling fragment in real production environment, Platform S1 which contains single GPU card on a node would be a good reflection of such circumstance. Since pipeline only parallelism might be an reasonable alternative under low-bandwidth interconnections, we also did PP-only weak scaling tests on GPT model. 

All of the SPMD-only and PP-only tests results are shown in Fig. \ref{fig:spmd_only}. It is proven that \OurSystem would help LM models getting high MFU in single node SPMD only parallelism, which is upto 68.1\% for GPT and 62.0\% for MoE on M8 platform. We can also see good scalability in more show cases in next subsection. All these results make the evidence that \OurSystem could easily enable a wide range of models from different application area.

From the strong scaling tests results of GPT-Medium model configuration, it could be seen that \OurSystem SPMD searching result also varies when the amount of running devices changed: For 2 GPUs, the strategy is as normally as DP. While for 4 GPUs, a variable in 'decoder layer 0' is made sharding among the devices and slightly data transfer decreasing is reported. When devices number becomes to 8, all 'attention Q/K/V' variables and there optimizer status variables are split out like DeepSpeed Zero-2 "$P_{os+g}$" \cite{deepspeed} schema, while others in the model are kept as DP. The searching report showed ~20\% data transfer size than DP. The running performance under such strategies supported the searching results on both Platform M8 and S1, compared with Alpa, which gave a more DP like plan on 8 GPUs.

\subsection{Generalizability on different models}

Besides GPT and GShard MoE models, we also involved other four models to prove generality, 1) VGG19\cite{vgg}, a
frequently adopted convolutional neural network for image recognition. We replace the model's last layer with a
million-scale classifier. 2) DNABert\cite{dnabert}, a scientific model for DNA sequences interpreting adopting with
Transformers and CNN. 3) Wide-ResNet\cite{wideresnet}, a variant of ResNet with larger channel sizes. We scale the model
size along with GPU numbers to study its scalability. 4) Unet, the key component in popular diffusion network. For
Wide-ResNet, we scale model size along with GPU numbers to study their FLOPS. For others, we scale batch size along with
GPU numbers to study their throughputs. With these models diversity, we could better investigate the generality of
\OurSystem. The experiment configuration lists in Table \ref{tab:batch_scale_conf} and Table \ref{tab:param_scale_conf}.

\begin{table}
\centering

\caption{VGG19 and DNABert benchmark configurations. \label{tab:batch_scale_conf}}
\resizebox{\columnwidth}{!}{
\begin{tabular}{lcccc}
\toprule
Model Name & $N_{params}$ & Batch Size / Device & $N_{GPUs}$\\
\midrule
VGG19 & 4.2B & 128 & from 1 to 8 \\
DNABert & 162M & 32 & from 1 to 8 \\
UNet & 33M & 16 & from 1 to 8 \\
\bottomrule
\end{tabular}}
\end{table}

\begin{table}
\centering
\caption{Wide-ResNet benchmark configurations. We keep global batch size at 4 for each experiment. \label{tab:param_scale_conf}}
\resizebox{\columnwidth}{!}{
\begin{tabular}{lccccc}
\toprule
Config Name & $N_{params}$ & $N_{channels}$& $N_{width_{factor}} $ & $N_{layers}$ & $N_{GPUs}$ \\
\midrule
WR-1.3B & 1.3B & 320 & 2 & 50 & 1 \\
WR-2.6B & 2.6B & 448 & 2 & 50 & 2 \\
WR-6B & 6B & 640 & 2 & 50 & 4 \\
WR-12B & 12B & 320 & 12 & 101 & 8 \\
\bottomrule
\end{tabular}}
\end{table}

All of the evaluation results are shown in Fig. \ref{fig:show_cases}. On one hand, both the scaling data experiment achieves near linear scalability. For VGG19, the classifer consumes huge device memory impedes training on single device. But it scales up with the suitable parallel strategy explored by \OurSystem. For smaller model like DNABert and UNet, \OurSystem prefers simple data parallelism. On the other hand, Wide-ResNet scales up on parameters which also achieves near-linear throughput.

\subsection{SPMD strategy searching performance}

In order to explain how one-pass SPMD strategy planning time is influenced by the complexity of the training model and
the end-to-end performance, we setup the experiment with another customized configuration set GPT model backbone (set
hidden size as 512, then change the layers for different cases), then run the SPMD-only parallel case 1-8 GPUs. Both
strategy planning time (on 2GPUs) and training step iteration time with data sample weak scaling was recorded. The whole
graph level searching (OPT\_LEVEL\_3, O3) and the sub-graph level searching (OPT\_LEVEL\_2, O2) were both tested in this
case. The results is shown in Fig. \ref{fig:opt_level}. 

The results indicate that O2 level method has better searching time then full graph searching (O3) when model layers is
more than 36 which has more than 30k HLO instructions in the graph. But the SPMD strategy given by O2 level method still
has ~10\% gap with O3, both from the engine report and from E2E performance on M8 platform. Besides the data transfer
size difference, the decrease in performance could also be attributed to the increased use of AlltoAll communication by
O2 search. However, such collective performance may not efficiently utilize the available communication bandwidth.

\begin{table}
\centering
\caption{The statistics of critical nodes and subgraphs in GPT-72-layers and MoE-32E-10B. \label{tab:critical_nodes_statistics}}
\resizebox{\columnwidth}{!}{
\begin{tabular}{lcccc}
\toprule
Model Name & instructions & critical nodes & proportion & subgraphs count \\
\midrule
GPT-72-Layers & 76k & 378 & 0.5\% & 377  \\
MoE-32E-10B & 18k & 90 & 0.5\% & 89 \\
\bottomrule
\end{tabular}}
\end{table}

The search time saving benefits from dividing the original graph into multiple subgraphs with coarser granularity. The
division depends on the identification of critical nodes. We collect the online statistics and do comparison with the
number of original operations, which shows that the proportion of critical nodes is quite small. This explicitly
contributes to the search time reduction especially significant for large models. Table
\ref{tab:critical_nodes_statistics} presents the statistics of two representative large models. It shows the critical
nodes proportion is around 0.5\% on both GPT-72-Layers and MoE-32E-10B. As a result, it reduces the original large graph
sizes from 76k and 18k to 377 and 89. This optimization greatly compresses the compilation time, while the runtime only
has a performance cost of 10\%, which is of great significance for actual production usage.

\section{Related Work\label{sec:related_work}}

\noindent\textbf{Data parallelism systems.} PyTorchDDP\cite{li2020pytorch} is the data parallelism training system works
on PyTorch, which synchronizes gradients using all-reduce. Distribution Strategy provides parameter server to train deep
models in asynchronous gradients descent for sparse models. Horovod\cite{horovod} designs the data
parallelism optimizer as the third-party plugin. It provides the automatic tuning approach by utilizing bayesian
optimizer to adjust the tensor communication group size. Zero\cite{zero} partitions state variables to improve memory
usage, which facilitates to train large models.

\noindent\textbf{Model parallelism systems.}
Mesh-TensorFlow\cite{mtf} provides SPMD style model parallelism interfaces to build models. \cite{gspmd} relies on the user annotations on HLO to get the SPMD plan. GPipe\cite{gpipe} divides deep models into stages and parallelizes them in pipeline manner. PipeDream\cite{pipedream} improves pipeline parallelism by using asynchronous training and integrates it with data parallelism. DAPPLE\cite{dapple} applies 1F1B schedule on synchronous training. TeraPipe\cite{li2021terapipe} discovers the pipeline parallelism for token level dimension on  transformer language models.

\noindent\textbf{Manually designed parallelization strategies.}
Megatron \cite{megatron} supports training Transformer models at large scale with expert-designed parallelization strategies that combines data parallelism and model parallelism. DeepSpeed\cite{deepspeed} optimizes kernel implementations besides expert-crafted parallelization strategies. DeepSpeed-MoE\cite{rajbhandari2022deepspeed} improves model architecture for MoE model and designs parallelizations on both the training and inference models. FasterMoE\cite{he2022fastermoe} introduces techniques to achieve dynamic load balance and optimize for congested all-to-all communication to accelerates MoE models training.

\noindent\textbf{Automated parallelism systems.}
Alpa \cite{alpa} formulates model parallelization in hierarchy, which models intra-operator parallelism to ILP problem and designs a dynamic programming method for exploring parallelism for intra- and inter-parallelisms. 
Unity \cite{unity} introduces a parallel computation graph and uses randomized graph substitution on the graph to jointly
optimize algebraic transformations and parallelization. FlexFlow\cite{flexflow} defines a "SOAP" search space and explores parallelization by using randomized search. Whale\cite{whale} partitions models guided by computation-balanced principle to accomodate
heterogenous compute devices, allowing specifying
parallelization strategies through parallelization primitives. Tofu\cite{tofu} minimizes
communication time in recursive search and automatically discovers parallelization
dimensions via interval analysis. PlaceTo\cite{placeto} uses reinforcement learning to decide device placement for model parallelism.

\noindent\textbf{Deep learning compilers.} MLIR\cite{lattner2020mlir} is a hierarchical and extensible compiler
infrastructure that standardizes the Static Single Assignment-based IR data structures. Relay\cite{roesch2019relay}
presents a compiler framework to unify and generalize IR in existing frameworks. TVM\cite{tvm} is a compiler
that exposes graph-level and operator-level optimizations to provide performance portability for DL workloads across
diverse hardware backends.

\section{Conclusion\label{sec:conclusion}}

We present \OurSystem, an industrial system that accelerates tensor programs with automatic parallelization. For
efficient and general SPMD strategy search, we design a three-level subgraph merging algorithm combined with the
formulated ILP and DP methods to find the solution. For pipeline parallelism, the stage division on non-linear model for
pipeline parallelism is formulated as an ILP problem and accelerated with a bound tightening heuristic. All the
heuristics are data-driven designed. They take critical part in reducing exploration search time thus pushing to the
general and automatic parallelization to industrial AI platforms. Finally, we implement a distributed runtime engine
with a task graph-based static scheduling that achieves barely observable runtime overhead.

\bibliography{main}
\bibliographystyle{plain}

\end{document}